\documentclass[fontsize=10pt]{article}
\usepackage[a4paper, margin=2cm]{geometry}
\usepackage{authblk}
\usepackage{siunitx}
\usepackage[version=4]{mhchem}
\usepackage{graphicx}
\usepackage[sort&compress]{natbib}
\usepackage{ulem}
\usepackage{multicol}
\usepackage{subcaption}
\usepackage{caption}
\usepackage{abstract}

\usepackage{todonotes} 
\usepackage{xcolor}
\usepackage[colorlinks]{hyperref}
\definecolor{DarkBlue}{RGB}{0,0,139} 
\hypersetup{citecolor=DarkBlue}
\hypersetup{linkcolor=DarkBlue}
\hypersetup{urlcolor=DarkBlue}
\usepackage{cleveref}
\title{Hair-thin confocal fluorescence endo-microscopy for deep-brain {\it in-vivo} imaging}
\newcommand*\samethanks[1][\value{footnote}]{\footnotemark[#1]}
\author[1]{\footnotesize Tom\'{a}\v{s} Pik\'alek\thanks{correspondence to tpikalek@isibrno.cz \& cizmart@isibrno.cz}}
\author[1]{Miroslav Stib\r{u}rek}
\author[1]{Tereza Tu\v{c}kov\'{a}}
\author[1]{Petra Kolb\'{a}bkov\'{a}}
\author[2]{Sergey Turtaev}
\author[3]{Jana Krej\v{c}\'{i}}
\author[1]{Petra Ondr\'{a}\v{c}kov\'{a}}
\author[1]{Hana Uhl\'{i}\v{r}ov\'{a}}
\author[1,2,4]{Tom\'{a}\v{s} \v{C}i\v{z}m\'{a}r\samethanks}

\affil[1]{\footnotesize Institute of Scientific Instruments of the CAS,
	Kr\'{a}lovopolsk\'{a}~147, 612~64 Brno, Czech Republic}
\affil[2]{Leibniz Institute of Photonic Technology,
	Albert-Einstein-Stra{\ss}e~9, 07745 Jena, Germany}
\affil[3]{Institute of Biophysics of the Czech Academy of Sciences, Kr\'{a}lovopolsk\'{a} 135, 612~65 Brno, Czech Republic}
\affil[4]{Institute of Applied Optics, Friedrich Schiller University Jena,
	Fr\"{o}belstieg~1, 07743 Jena, Germany}

\date{\vspace{-1.2cm}}
\begin{document}

            \maketitle
            \begin{abstract}
Confocal and multi-photon microscopy are widely used for {\it in-vivo} fluorescence imaging of biological tissues such as the brain, offering non-invasive access up to $\approx$\SI{1}{\milli\metre} depth without major loss in performance. A recently-developed alternative is holographic endoscopy, which exploits controlled light transport through hair-thin optical fibres. With minimal invasiveness, it provides observations at comparable spatial resolution, while extending its applicability to unprecedented depths. It has been used to resolve details of sub-cellular structural connectivity, record neuronal signalling, and monitor blood flow from the deepest locations of the living brain. Yet, its use, particularly in densely labelled brain regions, has so far been constrained by significant contrast loss, primarily due to the absence of a practical mechanism for rejecting out-of-focus fluorescence light -- a capability inherently provided by confocal and multi-photon microscopy.
Exploring opportunities in the structure of light modes of different MMF types we identify the possibility of achieving an analogue to confocal fluorescence microscopy through MMF-based endoscopes. Using a novel composite fibre probe that combines graded-index and step-index MMFs, we enable spatially resolved signal collection and selective rejection of out-of-focus light. This confocal filtering significantly enhances image contrast and resolution by suppressing background and off-plane signals. We demonstrate improved imaging performance on fine structural connectivity and intracellular calcium signalling in living mouse brain.
\end{abstract}
\begin{multicols}{2}
\section{Introduction}
Rapid advancements in the synthesis and analysis of structured light have
recently enabled full control over the light propagation through multimode
optical fibres (MMFs) and the generation of high-purity custom optical
fields emerging from the MMF\cite{Conkey2012OEa,DiLeonardo2011OE,Cizmar2011OE,Papadopoulos2013BOE,Ploschner2015NP,Morales-Delgado2015OEa,Gomes2022OE,Cizmar2012NC}. Among other applications\cite{Leite2018NP,Konstantinou2023OALIE}, this prospect has
shown immense potential for {\it in-vivo} neuroscience imaging, introducing
hair-thin endoscopes which enable high-resolution, single-photon imaging of
fluorescently labelled cells and blood vessels deep within the brain\cite{Ohayon2018BOE,Turtaev2018LSA,Stiburek2023NC}. This
approach facilitates extremely atraumatic and detailed {\it in-vivo} examination
of deep-brain structures when compared to the state-of-the-art
instrumentation based predominantly on the exploitation of GRIN lenses\cite{Flusberg2005OL,Ghosh2011NM,Barretto2011NM,zong2022large,d2022compartmentalized}.

The method treats the fibre as an optically complex, or random medium, where
light transport remains deterministic and can be described by a linear
operator known as the transmission matrix (TM)\cite{Popoff2010PRL,Cizmar2011OE,Conkey2012OE,Rothe2019AS}. Holographic wavefront
shaping is employed to overcome this complexity: Utilising the empirically acquired TM and a suitable computer-controlled spatial modulator of light, a sequence
of diffraction-limited foci is formed at the distal end of the MMF,
effectively scanning the scene point by point. After interaction with the sample, returning signals are partially
captured by the MMF within its spatial and spatial frequency constraints.
In the reflectance imaging regime, wavefront shaping can also be used to process the returning highly coherent light signals, allowing amongst other benefits to mimic confocal imaging\cite{Loterie2015OL}. 
In the fluorescence regime, the returning signals are generated via the single photon process and therefore they lack temporal coherence. Hence, the holographic wavefront-shaping technologies and methods can no longer be used to unscramble these returning signals.
Instead, they are detected at the fibre's
proximal end, using a bucket detector, disregarding their spatial distribution.
Each pixel of the resulting image therefore contains signal contributions generated along the
whole extent of the excitation beam, with no option to
discriminate whether or not they originated from the
high-intensity region of the focus. Resulting images then suffer from loss of contrast and strong contamination by out-of focus features.

In practical terms of neuroscience imaging, this makes the identification of
fine details of structural connectivity as well as signalling activity in densely labeled regions challenging, restricting the use of
the technology to sparsely labelled samples.

MMFs share several similarities with other random media. But there are also
very important cases, where they stand out. The negligible transmission
losses and the near-perfect cylindrical symmetry result in very robust
input-output correlations, which can be exploited even if the propagating
signal lacks coherence. In this work we exploit such special properties of
both, the step-index (SI) and the graded-index (GI) MMFs in order to empower
the MMF-based holographic endoscope with the possibility of out-of-focus
light rejection.

Consider a plane wave coupled into a step-index multimode fibre (SI MMF) at an angle relative to the fiber axis, within the acceptance range defined by the fibre's numerical aperture (NA). Such wave would only excite modes
with closely similar propagation constants, matching the projection of the
wave’s $k$-vector on the fibre’s axis. When these modes reach the distal
extremity of the SI MMF, they enter the free space behind as a spectrum of
plane-waves with the axial component of their $k$-vectors closely matching
that of the input field. The far-filed of such outputs thereby features a
narrow annulus. This leads to a very robust input-output correlation, posing
that light signal originating from certain annular zone of the input far
field will be found at the corresponding annular zone of the output far
field. The complexity of light transport is therefore reduced to ‘scambling’
within each of these zones (azimuthally), but the light doesn't experience any
significant mixing amongst the zones (radially). This effect has been shown
to sustain extremely well over several meters even in considerably curved
MMFs\cite{Cizmar2012NC} and emerges regardless of the state of the light’s temporal coherence.

The existence of this robust input-output correlation across the distal end's far field is, on its own, 
not directly exploitable in fluorescence imaging. Its potential would appear  
if it was instead available in close proximity to the MMF's distal extremity, across a desired focal plane.
This is when the GI MMFs,
featuring the self-imaging effect, come into play. Due to the specific
structure of the GI MMF modes, which meet with almost the same phase at
regular intervals along the fibre length (the pitch length), these class of
MMFs periodically reforms the input field along the propagation. At half
the pitch length, the input field appears again rotated by 180° and the
quarter pitch essentially works as a lens forming the required conversion to
(and from) the far field (Fourier transform). Splicing a quarter-pitch
length of a GI MMF (the GI endcap) on the distal end of an arbitrary long SI
MMF one therefore obtains a probe in which the input far field annular zones
connect to the corresponding annular zones directly at the distal facet.
Using slightly shorter GI endcaps, it is possible to shift the plane, across
which this feature peaks, away from the facet, thereby providing a specific
working distance. Interestingly, this  function remains robust even when the probe is equipped with the side-view termination \cite{Silveira2021OE}, transposing this significant plane off-axis. This enables imaging tissues unaffected by the probe insertion and providing continuous atlas view across unprecedented depths\cite{Stiburek2023NC}.   
\begin{figure*}[ht!]
	\centering
	\includegraphics[width=\textwidth]{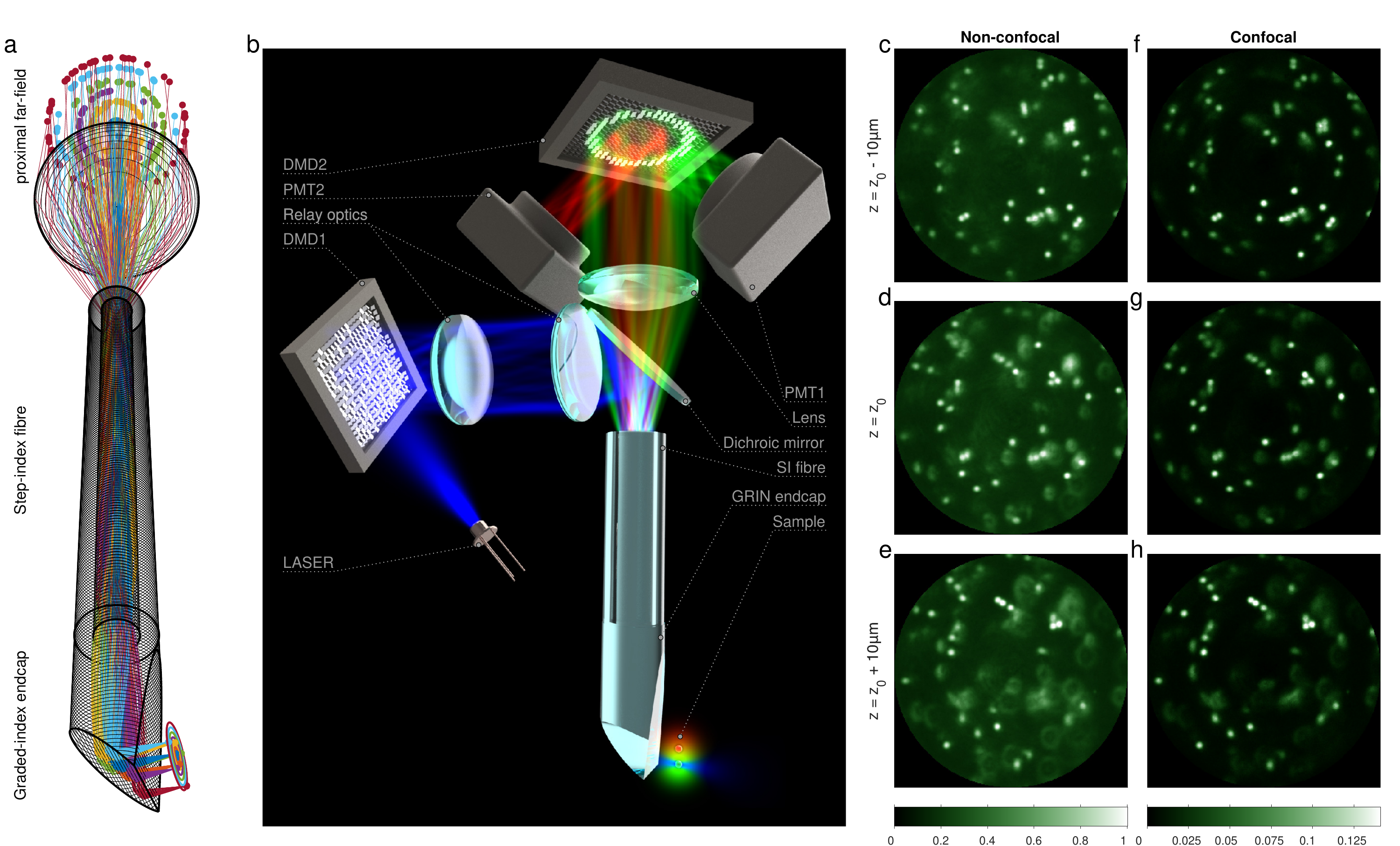}
	\caption{{\bf The concept of confocal endo-microscopy.} {\bf a,} Ray model of the light propagation through side-view MMF fibre probe. Light from a point at the focal plane ($z=z_0$) is converted by the graded-index (GRIN) fibre endcap to a collimated beam, which is, due to the conservation of the propagation constant, delivered by the step-index (SI) fibre to an annular zone of the proximal far-field. {\bf b,} Simplified concept of the imaging geometry. DMD1 provides structured excitation light resulting in a diffraction-limited focus scanning along the distal end focal plane. Fluorescent signal is collected backwards through the probe, isolated from the excitation path and spatially filtered at DMD2. Confocal images are formed by signal detected at PMT1, non-confocal images are formed by combining signals from both PMT1 and PMT2.  {\bf c-e} and {\bf f-h,} Series of images of fluorescent beads (\SI{2}{\micro\metre} in diameter) obtained in the non-confocal and the confocal regime respectively.  For all confocal images, the confocal parameter (width of the virtual aperture annulus) has been set to $w/w_{floor}=1$. Images {\bf d}  and {\bf g} were taken with the empirically obtained TM.  {\bf c}  and {\bf f} were obtain with numerically recalculated TM for a focal plane shifted by \SI{10}{\micro\metre} towards the fibre probe. In contrast, {\bf e}  and {\bf h} were obtain with a TM, numerically recalculated for a focal plane shifted \SI{10}{\micro\metre} away from the fibre probe. All images {\bf c}--{\bf h} are shown with no alpha manipulation. The minimum of the false-colour interval corresponds to the true (calibrated) zero of the PMT detectors, the upper boundary is set such, that the maximum of \SI{0.1}{\percent} of pixels is saturated. The colour-bar for the confocal modality is calibrated relatively to the non-confocal counterpart to accurately reflect the signal loss due to the confocal rejection.     
	}
	\label{fig:principle}
\end{figure*}
Exploiting the GI-endcapped probe for out-of-focus light rejection
necessitates no modifications to the excitation path (provided the light modulator lies in plane conjugate to the proximal facet of the MMF probe\cite{Leite2021AP}). 
 In the detection path,
we however need a dynamic annular aperture across the distal farfield plane,
which changes its radius proportionally to the distance of the excitation
focus from the fibre axis, in synchronous manner with the scanning sequence.
The fluorescent signal originating from the sample point is thereby filtered
by the annular function rejecting light from other locations.

In this paper we describe the considerations of obtaining the optimal probe,
our technological solution for the dynamic aperture and all necessary
calibration and optimisation procedures to maximise the utility of this
prospect for {\it in-vivo} fluorescence imaging. Further, we employ this technology to provide {\it in-vivo}
observations inside deep brain regions of various mouse models and showcase its benefits
for the structural connectivity observations as well as acquisition of the
signalling activity.

\section{Results}
\subsection{The optical system and imaging procedures}
The light transport mechanism behind the confocal imaging, which is enabled by the composite probe is outlined in Fig. {\bf \ref{fig:principle}}. Panel {\bf \ref{fig:principle}a} represents an illustrative ray-model of the light transport for the side-view terminated composite MMF fibre probe, featuring mirror facet under \SI{45}{\degree} with respect to the fibre axis as well as a transmissive front facet under \SI{5}{\degree}. 
The medium outside the distal end of the probe (the graded-index fibre termination) is considered to be water (refractive index of 1.33), while the medium outside the proximal end (flat step-index termination) is considered to be the air (refractive index of 1). 

Individual light paths originate outside the probe near its distal extremity in a series of foci. Each focus lies in the focal plane for which the length of the endcap has been optimised. Each focus is further located at a specific radial distance from the central position, where the optical axis of the fibre would be reflected. The model traces the rays all the way to the proximal far-field plane, passing through the water-filled space, the side-view-terminated GRIN endcap, the SI segment and the proximal lens surrounded by two air-filled volumes. The model verifies that the radial positions of the light-paths across the initial and the final plane are bound by a simple linear relationship. The analogous ray model for the traditional (straight-view) configuration and the equivalent wave model of the light transport through the probe is available in the Supplementary information (Figures \ref{ray_str}-\ref{wave_gfp} and Media \ref{wave_lambdas}-\ref{wave_movcirc}). 

The side-view optical probes used throughout this study are in-house manufactured from commercially available SI and GRIN MMFs.  
The spatial resolution of imaging is for the composite probe constrained by both, the NA of the GRIN MMF and the core size of the SI MMF, while the field of view is constrained by, the core size of the GRIN and the NA of the SI MMF. The specifications have therefore been chosen to impose similar constraints, while trading-off the tissue damage and performance: both types feature a core size of \SI{100}{\micro\metre} and numerical apertures of 0.22 and 0.29 have been selected for the SI and the GRIN MMFs, respectively. This leads to the achievable resolution of about one micrometer within the central region of the FOV, yet towards the edges of the FOV the resolving power would gradually drop, due to the spatially varying mode density of the GRIN fibres.   
Protocols for manufacturing the fibre probes are explained in Methods \ref{met:probe_prep}. 

The optical system used in this study is derived from \cite{Stiburek2023NC} and it maintains all its features. It is described in full detail in Methods \ref{met:system} together with all necessary calibration procedures. 
Fig. {\bf \ref{fig:principle}b} illustrates the main constituents of the experimental geometry for simpler understanding of its functions. Once the calibration of the excitation path (shown in blue) is completed, the empirically acquired TM is used to produce diffraction limited foci across the focal plane. This is achieved by appropriate setting of the digital micro-mirror device (DMD1) providing the necessary light modulation. For each of their mirror elements, DMDs only offer the choice between two discrete orientations. Yet, this is sufficient for achieving highly complex vector field distributions \cite{Mitchell2016OE,Turtaev2017OE,Gomes2022OE}. Importantly, it provides the highest modulation rates (exceeding \SI{30}{\kilo\hertz}) from the commercially available technologies, making it at present the optimum choice for our desires. 
\begin{figure*}[htb!]
	\centering
	\includegraphics[width=\textwidth]{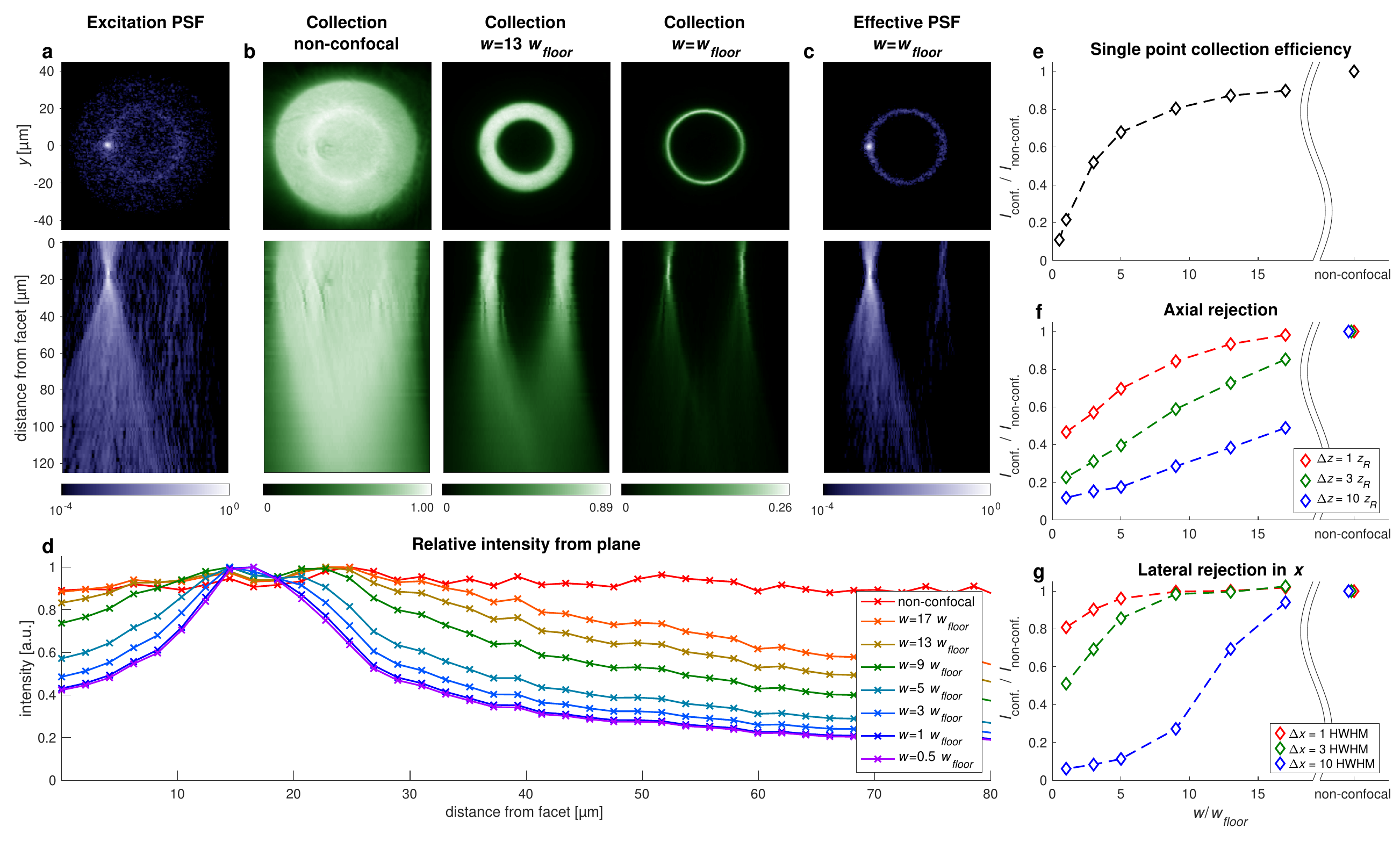}
	\caption{{\bf Quantitative assessment of out-of-focus signal suppression.}
		\textbf{a}, Lateral and axial point spread function (logarithmic scale).
		\textbf{b}, Collection PSF for a non-confocal regime and two confocal
			factors $w$ (linear scale).
		\textbf{c}, Effective PSF calculated as a product of the excitation and
			collection PSFs (logarithmic scale).
		\textbf{d}, Measured average collection efficiency as a function of the 
			distance from the facet for non-confocal and confocal regimes with
			different confocal factors.
		\textbf{e}, Measured collection efficiency at the focal point, normalised 
			to that of the non-confocal regime, as a function of the confocal factor.
		\textbf{f} and \textbf{g}, Measured collection efficiency, normalised to the 
			non-confocal regime, for focal points displaced axially (\textbf{f}, 
			$\Delta z$, where $z_R$ is the Rayleigh length of the excitation focus) 
			and laterally (\textbf{g}, $\Delta r$, measured in units of the 
			half-width at half-maximum of the excitation focus).
		}
	\label{fig:psf}
\end{figure*}

The quality of the resulting diffraction limited foci is quantified by a metric of power ratio (PR), expressing the fraction of the optical power carried by the focus with respect to the overall power delivered by the probe. There are numerous factors affecting the PR\cite{Stiburek2023NC} and the optical system reached values around 70\%, when the phase-only modulation regime has been applied\cite{Gomes2022OE}. Note that in the excitation optical path, the DMD1 lies in the conjugate plane of the proximal probe facet. Should the Fourier (far-field) plane be used instead, akin to many traditional wavefront-shaping systems, the nature of the composite probe would cause that only light from a narrow annular zone of the DMD would be utilised in forming the focus, while the remaining signal would manifest itself as further undesired contamination of the focal plane\cite{Leite2021AP,Stellinga2021S}, leading to significantly lower values of PR. For purely step-index fibre probes\cite{Stiburek2023NC}, the remaining undesired power is distributed uniformly across the whole field of view as a random speckle pattern. For the confocal composite probe, the remaining signal concentrates predominantly across the annular zone, at which the desired focus lies (see Fig. {\bf \ref{fig:psf}a}).  

When applied for imaging, the fluorescence signals originating from the focal region (shown {\bf in Fig. \ref{fig:principle}b} in green) are collected backwards and projected by the composite probe and the proximal lens onto the plane of the DMD2 located in the fluorescence-harvesting path in the proximal far-field plane. DMD2 constitutes the hardware solution of the dynamic aperture. The mirrors located in the annular zone which correspond to the actual radial position of the focus are oriented such that they reflect the signal onto the bucket detector PMT1. Light originating outside the focal volume (illustrated in red) is in analogous manner delivered onto the DMD, where it however meets mirrors set to the complementary orientation. These signals can either be rejected or detected by separate bucket detector PMT2, for the purpose of the system's performance evaluation (mimicking the standard, non-confocal regime).      

Initial qualitative demonstration of the system is provided in Figs. {\bf \ref{fig:principle} c}-{\bf h}. Confocal regime (signal detected at PMT1) is compared to the non-confocal (signals from both detectors are combined, taking into account their varying response), showing significant improvement in the contrast. This study also includes the cases where the TM has been numerically recalculated, shifting the output plane into different distances away from the fibre facet \cite{Ploschner2015NP,Stiburek2023NC}, in front and behind the optimal confocal plane. The results show that the confocal ability is well-preserved across several 10s of \si{\micro\metre}.    

\subsection{Performance assessment}
To quantify the performance of the confocal imaging system through the multimode
fibre, we characterised the spatial distribution of the confocal point spread
function (PSF), as summarised in Fig. \ref{fig:psf} (see Methods for the
detailed protocol). The excitation PSF, shown in Fig. \ref{fig:psf}a,
illustrates the diffraction-limited focal intensity distribution generated at
the fibre distal end for a selected focal point position. The corresponding
collection PSFs for the non-confocal configuration and under varying confocal 
factor $w$  are presented in Fig. \ref{fig:psf}b. 
 The confocal factor $w$ represents the thickness of the ring-shaped confocal aperture on DMD2. We express $w$ relative to $w_{floor}$ -- the expected lower bound below which the axial confocal function is expected to plateau. Throughout this work, $w_{floor}$ is set to \SI{61.8}{\micro\metre} (see Methods for details). 
%
 As the factor $w$ reduces, the collection PSF becomes
progressively narrower both axially and radially. Due to the cylindrical
symmetry of the collection function, no filtering occurs in the tangential
direction. The effective PSF, obtained as the product of the excitation and
collection functions (Fig. \ref{fig:psf}c), therefore exhibits increased
confinement around the focal plane for smaller confocal factors, indicating
enhanced optical sectioning and improved suppression of the fluorescence
generated by the speckled background around the excitation focus. This behaviour
is further reflected in the measured collection efficiency profiles (Fig.
\ref{fig:psf}d–g). Decreasing $w$ results in a reduction of the out-of-focus
fluorescence intensity by up to a factor of five compared to the non-confocal
regime. However, when the confocal factor is reduced below approximately $3w_{floor}$, the overall collection
efficiency drops steeply while the improvement in optical filtering becomes
marginal. 
These results collectively confirm that confocal detection through a
multimode fibre significantly enhances spatial selectivity, providing effective
rejection of out-of-focus and background fluorescence without altering the
excitation conditions.

\subsection{Applications in  {\it \textbf{in-vivo}} imaging}
\begin{figure*}[!ht]
	\centering
	\includegraphics[width=\textwidth]{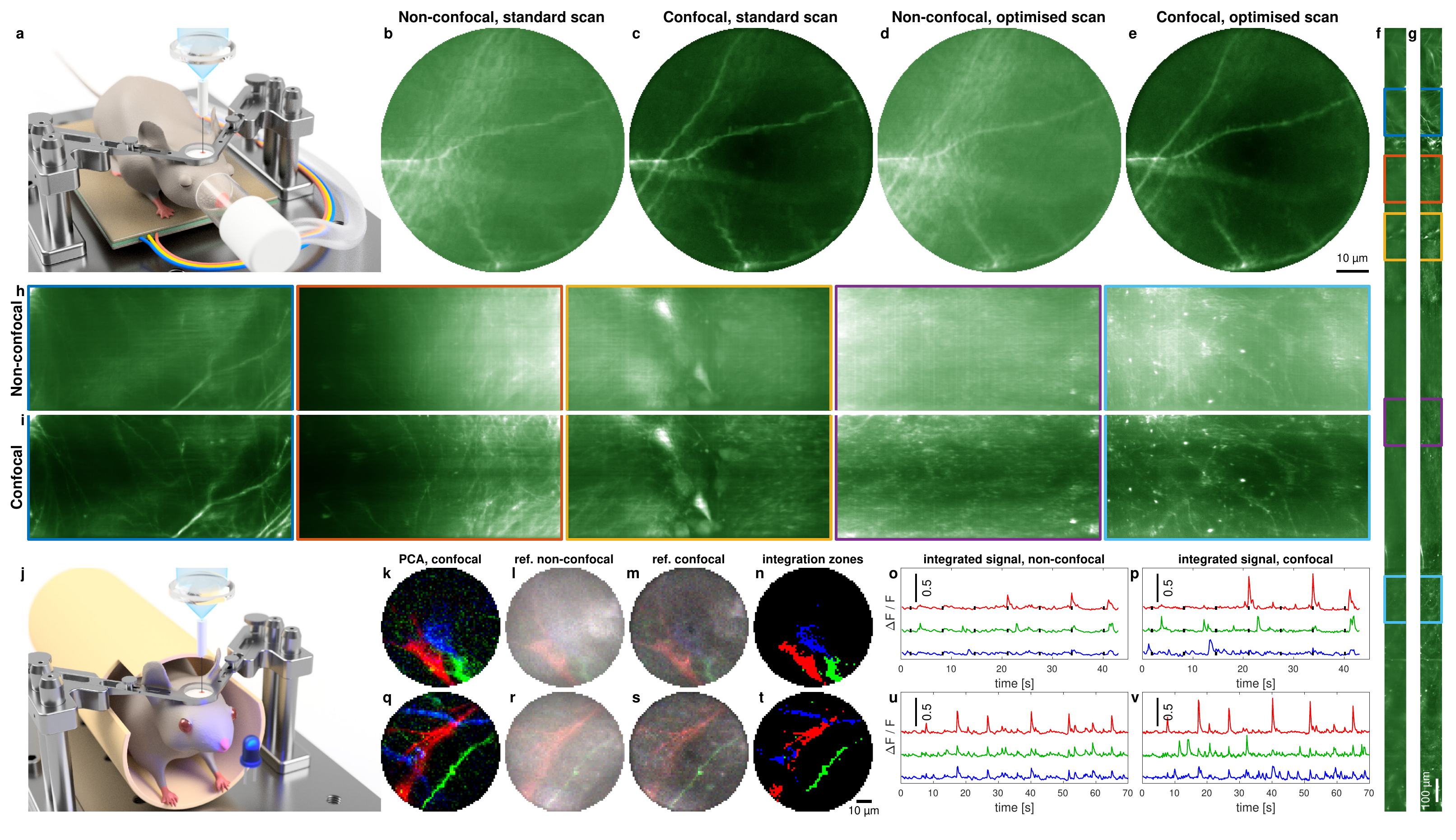}		
	\caption{
		{\bf Confocal endo-microscopy in deep-brain structures}.
		\textbf{a}, Simplified experimental arrangement of structural connectivity imaging in anaesthetized animal model.  
		\textbf{b}, \textbf{d}  and \textbf{c}, \textbf{e}, Full-field of view non-confocal and confocal records of 
		neurons  
		in a Thy1-GFP-M line. 
		\textbf{b} and \textbf{c} are obtained using the standard (raster) scanning sequence, while \textbf{d} and \textbf{e} are obtained with the `block-wise' scanning sequence designed to suppress the heart-beat-induced motion artefacts. 
		\textbf{f} and \textbf{g}, Atlas-view non-confocal and confocal records across the brain depth region spanning \SI{4.1}{\milli\metre}. To visualise whole extent of the brain in one image, the brightness is locally scaled.  
\textbf{h} and \textbf{i}, Full-resolution samples from \textbf{f} and \textbf{g} respectively. 
Panels \textbf{f} -- \textbf{e}, \textbf{h} and \textbf{i} are shown with no alpha manipulation. The minimum of the false-colour interval corresponds to the true (calibrated) zero of the PMT detectors, the upper boundary is set such, that the maximum of \SI{0.1}{\percent} of pixels in each image is saturated.
		\textbf{j}, Simplified experimental arrangement of signalling activity monitoring in awake animal model. 
		\textbf{k--p}, Monitoring GCaMP activity of cell somata in a Ai162D$\times$Camk2a-CreERT2 mouse. \textbf{k}, Three strongest principal components of the record shown in RGB channels. \textbf{l} and \textbf{m}, Reference frames with the highest score of the three principal components obtained in the non-confocal and the confocal modality, respectively. \textbf{n}, Selected integration zones for activity monitoring. 
		\textbf{o} and \textbf{p}, Recorded fluorescence signal collected within the integration zones, in the standard $\Delta F / F$ form, i.e. the relative change in fluorescence intensity over the baseline. 
		During the record, the animal model has been exposed to a visual stimulus indicated as black dashes.  
		\textbf{q--v}, The equivalent of \textbf{k--p} for the records of neuronal processes (dendrites). No visual stimulus has been applied.  In all studies, the confocal parameter (width of the virtual aperture annulus) has been set to $w/w_{floor}=3$. 
	}
	\label{fig:imaging}
\end{figure*}

To demonstrate the enhanced performance of the confocal modality in {\it in vivo} observations, we focus on the two most common regimes -- observations of the structural connectivity and the signalling activity. The structural connectivity has been performed in sedated animal models expressing fluorescent proteins (see Fig. \ref{fig:imaging}\textbf{a} for simplified arrangement of the experiment). 

Panels {\bf b -- e} of Fig. \ref{fig:imaging} are full field of view images comparing the non-confocal and the confocal images of 
neuronal 
dendrites (Thy1-GFP-H) in hippocampus (see also Supplementary Media \ref{hb_demo} for consecutive sequence of recorded frames and Supplementary Figure \ref{full_field_other} for images from other animal models).
Due to the heart beat, the tissue exhibits periodic motion with respect to the probe, with amplitude of a few \si{\micro\metre} and a frequency of $\approx$\SI{10}{\hertz}, remaining very stable under anaesthesia. Standard scanning sequence (row by row) results in strong motion artefacts visible in panels {\bf b} and {\bf c}. These appear since the neighbouring rows are taken at different phases of the heart-beat cycle. This makes identification of fine structures (e.g. dendritic spines) challenging as their images are significantly stretched or compressed. This issue can be suppressed by applying optimised scanning sequence of heart-beat-long segments (see panels {\bf d} and {\bf e}). Each of such segments, containing a certain amount of pixel rows, is scanned column-wise, thereby minimising the heart-beat phase difference amongst neighbouring pixels. Analogous optimised scanning sequences are also available for the volumetric and the atlas-view regimes discussed below (see Methods \ref{met:HB compensation} for more details).       

%
%

Continuous descent of the side-view fibre probe enables completing the `atlas view' -- uninterrupted record along the whole depth of the brain\cite{Stiburek2023NC}. Complete atlas view in the non-confocal and the confocal modalities are shown in panels {\bf f} and {\bf g} respectively, with three selected regions shown in full resolution  in { \bf h} and { \bf i}. 
The dataset is complemented by Supplementary Media \ref{atlas_orig_gif} -- \ref{atlas_enh_mp4}. Here we show the full-resolution volumetric records (showing 10 focal planes separated by \SI{2}{\micro\metre}).
While Fig. \ref{fig:imaging} preserves the data as genuinely as possible without any contrast-enhancement,  Supplementary Media \ref{atlas_enh_gif} and \ref{atlas_enh_mp4} offer also the opposite, where we locally apply the same contrast and gamma-enhancement filters on both confocal and non-confocal records. 
The above examples show that the confocal modality allows significant improvement in contrast, and the possibility to identify structures which are irresolvable in the standard, non-confocal modality.

Whereas structural connectivity remains unaffected whether the animal is sedated or fully awake, signalling activity is heavily influenced by anaesthesia.
Therefore, we have introduced a number of changes in the animal handling protocol (see Methods \ref{met:Surgery}) as well as the animal housing apparatus (see Fig. \ref{fig:imaging}\textbf{j} for simplified arrangement), enabling observations of the signalling activity in awake animals. To visualise the activity, we have exploited fluorescence protein GCaMP6s, expressed in neurones and their processes in a Ai162D$\times$Camk2a-CreERT2 mouse line (Fig. \ref{fig:imaging}{\bf k} -- {\bf p}).
In order to cover the whole FOV at frame-rates sufficient for capturing the signalling activity ($\approx$8 frames per second), we have decimated the pixel grid by factor of 4 in both directions leading to the resulting pixel separation of \SI{1.14}{\micro\metre}, still only slightly larger that the diffraction limit.    
The gallery of full FOV frames was acquired simultaneously in both, confocal and non-confocal modalities focusing on neuronal somata as well as their processes (see Supplementary Media \ref{sig_soma} and \ref{sig_proc}). While for the somata we have applied a visual stimulus, the activity recorded at the processes is spontaneous. The confocal modality results have been subjected to PCA analysis (see Fig. \ref{fig:imaging} {\bf k} and {\bf q} for the strongest components and {\bf l}, {\bf m}, {\bf r} and {\bf s} for corresponding highest-scoring frames), in order to identify zones across the FOV from which the fluorescence signal has been integrated (panels {\bf n}, {\bf t}). In both studies, the integrated signal for the confocal modality (see Fig. \ref{fig:imaging} {\bf p} and {\bf v}) features much higher deviation from the base-line, due to the suppression of the undesired out-of-focus signal contributions.

\section{Discussion}
In this paper, we have shown that the combined input-output correlations of light transport through the step-index and the graded-index multimode fibres provide a route to determine the radial position of an incoherent light source from the transmitted light. The same correlations have important implications also in the time domain. When forming a focus through the composite MMF probe in the pulsed regime, all spectral components propagate through the SI part of the probe at the same angular mode group and therefore experience significantly reduced mode dispersion. This then opens the route towards other advanced methods of modern microscopy relying on short pulses (e.g., multiphoton and pulsed super-resolution microscopy\cite{gomesSTED2025_Arxiv}).  
 Together with the significance for the spatial domain, this can lead to many important enhancements in applications relying on controlled light transport through multimode fibres. For example in imaging applications, one must no longer use the DMD to form each focus individually, one at the time. Instead one may produce a number of foci to illuminate different points across the FOV, each at different radial distances from the axis, and separate their fluorescent responses at the proximal far-field plane by series of concentric annular detectors. This can lead to acceleration of image acquisition in excess of one order of magnitude. 

We have applied these principles to achieve the equivalent of confocal fluorescent imaging, although the light rejection is only applicable along the radial direction. Our parametric studies have confirmed significant  suppression of out-of-focus light in both the lateral and the axial directions, leading to enhanced contrast as well as improved sectioning ability. Our in-vivo demonstrations verify the prospect's applicability to deep-brain imaging, bringing about remarkably  improved access to observe fine deep-brain structures and monitor signalling activity of neuronal cells and their processes. 

The confocal feature is tuneable by the settings of the virtual aperture ($w$ factor), allowing one to trade off between image quality (contrast) and strength of the signal (particularly its contamination by shot noise). This would be particularly important in photon-hungry applications such as the voltage imaging. 

An important topic to discus is the calibration procedure. The hardware part of the excitation path calibration (acquisition of interferograms for TM computation) utilises the fastest commercially available technologies and can be currently achieved in a few minutes. Further, through the implementation of highly parallelised and efficient computer coding the TM processing and computation of the DMD binary modulations can also be achieved in a few minutes\cite{Stiburek2023NC}.  In the confocal regime, we have yet another calibration procedure, which links the position of the excitation foci to the centre and the radii of the virtual aperture across the proximal far-filed plane. Our implementation (see Methods) has been designed mainly to verify this concept with high accuracy and with minimal modifications to the existing optical path, rather than for its completion in shortest possible time. We already foresee non-elaborate routes to accelerate this procedure significantly for example with the employment of a LED array or display, rather than using the single incoherent light source anchored to the positioning stage. Therefore, even if all calibration procedures have to be executed before imaging can commence, for example when the confocal fibre probe is to be exchanged for a new one, this wouldn't cause a significant delay in experimental work beyond about \SI{10}{\minute}.      

Due to limited collection efficiency, the photon budget remains the bottle-neck of the MMF-based holographic endoscopy discipline in general, with possible routes forward through the enhancement of the MMF numerical aperture (GRIN MMF for the confocal modality). While in this study our choice reached the maximum NA across commercially available products, we are currently exploring the possibility to significantly enhance this quantity in custom-made fibres, which will be reflected in our future work.  

State-of-the-art microscopes and endoscopes used in {\it in-vivo} observations are, to a large extent, tolerant to small mechanical and thermal drift of the geometry. This is not the case for MMF-based endoscopes, since small drifts in the order of the diffraction limit can lead to complete loss of the imaging function. Therefore robust and temperature controlled optical paths are required to maintain the quality of imaging during the experimental sessions spanning typically over several hours. These drift-associated problems have been solved through a decade of intensive research and the standard non-confocal MMF-based endoscope has already reached the technology readiness allowing for its transfer. Since relatively small changes in the optical geometry are required to enable the confocal modality, it is likely that this powerful prospect will be made broadly available in the upcoming years, assisting neuroscience in the exploration of the brain functions and in its quest to identify combat strategies against severe neuronal disorders. 

 \section*{Acknowledgements}

We would like to acknowledge support from the European Research Council
(724530, 101069245, 101082088 and 101158010), the Ministry of Education, Youth and Sports of the Czech Republic
(CZ.02.1.01/0.0/0.0/15\_003/0000476 and CZ.02.1.01/0.0/0.0/16\_013/0001775), the
European Regional Development Fund (LM2018129), the European Union’s H2020-RIA
(101016787), Czech Technology Agency - Centre of Advanced Electron and Photonic Optics (TN02000020) and the Institute of Scientific Instruments of the CAS. We acknowledge the Czech-BioImaging facility of the Institute of Scientific Instruments CAS, ISI-MR LM2023050 for hosting our animal experiments. We would like to express our gratitude to Zenon Star\v{c}uk jr. and his team for sharing their facilities,
assistance with animal models and advice.
CzechNanoLab project LM2023051 funded by MEYS CR is acknowledged for the financial support of the parylene layer fabrication at CEITEC Nano Research Infrastructure.
Further we would like to express our gratitude to Jan Pavelka for support in development of custom head-plate and stereotactic navigation system, Jan Novotn\'{y} for support in developing the custom cover-glass solutions featuring access ports and to Andre Gom\'{e}s for assistance in software solutions and proofreading the manuscript.

\section*{Author contributions}

T.\v{C}. conceived the background for the presented technology.
M.S., T.P., H.U. and T.\v{C}. build the experimental geometry.
T.P., M.S., and S.T. compiled the computer controlling interface.
M.S., T.P., and P.K. developed and manufactured the custom fibre probes.
J.K. and P.K. developed and maintained the used animal lines.
P.O. and T.T. developed animal and surgical protocols.
H.U. and P.K. performed surgeries.
T.P. and M.S. performed all imaging and probe characterisation experiments.
T.P. and T.\v{C}. analysed the data.
T.\v{C}. and H.U secured funding 
T.\v{C}. led the project.
T.\v{C}., T.P. and H.U. wrote the manuscript with contributions from all
authors.

\section*{Competing interests}
T. \v{C}. and S. T. are founders and share-holders of a limited liability company DeepEn GmBH, which specialises on transfer of MMF-based endoscopes. T. \v{C}. and S. T., T. P. and M. S. are authors of a related patent DE102024202794B3,  ``Endoscopic detector system, use of a composite optical fiber, endoscopic system and method for examining a sample.'' 

\section{Methods}

\subsection{Experimental system}
\label{met:system}

In order to filter the fluorescent signal collected through the MMF, the system operated two DMDs in synchronous manner. The first DMD (DMD1) modulated the illumination signal as detailed elsewhere \cite{Stiburek2023NC} together with the calibration procedures of the excitation signal modulation and protocols for non-confocal {\it in-vivo} deep-brain imaging. 
The second DMD (DMD2), was deployed in the detection unit to process the returning fluorescent signals spectrally separated from the illumination path. 
It was placed in the far-field plane of the proximal end of the probe to provide the dynamic spatial filtration of the collected light. 

The symmetric
orientation of the DMD mirrors' `on' and `off' states allowed for separating the fluorescence signals, and directing them onto two
photomultiplier tubes (PMTs).  This way, 
the light accepted by the mask (confocal imaging) and the rejected light can be acquired simultaneously. 
Combining both signals produced a non-confocal image (an image without
filtration of the collected light).

The modifications made to the calibration module involved using a
non-polarizing beam splitter to split the beam into two paths: one for
excitation beam calibration and another for confocal detection calibration.
To calibrate the excitation beam, the focal plane behind the distal end of the composite fibre probe was imaged onto a camera and overlapped with a reference beam. For
confocal detection calibration, the same plane was focused onto a
single-mode fiber (SMF). This fiber served as a spatial filter and played a
dual role during confocal calibration. A photodiode at the other end of the
fiber was utilized to measure the intensity of the excitation beam coupled
into the SMF, allowing for the measurement of the relative position of the
SMF and the endoscope's field of view. Additionally, a green LED was coupled
into the SMF, creating a point source in the focal plane of the MMF.

\subsection{Probe preparation}
\label{met:probe_prep}
The confocal probe was constructed by splicing a Thorlabs FG105LCA step-index
fiber (\qty{100}{\um} / \qty{125}{\um} core / cladding diameter, \qty{0.22}{NA}) and
a DrakaElite graded-index fiber (\qty{100}{\um}/\qty{140}{\um}, \qty{0.29}{NA})
at the distal end. This fiber combination produced a field of view with a
diameter of \qty{75}{\um}, and the numerical aperture of the probe was~0.29.

The graded-index fiber was fusion-spliced (using 3SAE Large Diameter Splicing
System LDS 2.5) onto a roughly \qty{25}{\mm} long segment of the step-index fiber,
which was right-angle cleaved at both ends. The graded-index fiber was also
right-angle cleaved to a length slightly longer than a quarter pitch. The
resulting probe was affixed into a ceramic ferrule. The splice at the distal end
of the probe underwent polishing at \qty{45}{\degree} and \qty{5}{\degree} using
a custom-built polishing system. Additionally, a thin layer of aluminum was
deposited on the \qty{45}{\degree} face using a vacuum evaporator (JEOL JEE-420)
to create a side-view probe, as reported in \cite{Silveira2021OE}. This
polishing step was used for adjusting the splice length to achieve probes with a
working distance of about \qty{15}{\um}. Probes used for {\it in-vivo} imaging were
further coated with a thin parylene layer using SCS PDS 2010.

\subsection{Calibration procedures}

For successful calibration of the system, it is critical to ensure that the confocal plane of the probe is sharply imaged onto the camera inside the calibration module (components marked by an asterisk in Supplementary Figure \ref{suppl:fig:setup} are rigidly mounted on a single 3D positioning stage, enabling computer-controlled alignment).
As an initial estimate of the optimal position (alignment of the calibration module with respect to the fibre probe), the DMD1 shapes the excitation signal being coupled into the MMF so it carries only a single spatial frequency, i.e.,  entering the proximal fibre facet at an angle close to the numerical aperture limit. This signals excite a single SI-fibre mode group and forms a ring across the focal plane of the fibre probe. The plane corresponding to the minimum thickness of the output ring is then brought into focus on the camera. To refine the alignment more precisely, we employ a fluorescent phantom -- an incoherent signal propagating through the calibration module in the direction opposite to the excitation path. This signal forms a focus in the focal plane of the calibration objective and, provided the alignment is correct, lies within the confocal plane of the fibre probe.
The fluorescent phantom is realised using a green LED coupled into a single-mode fibre (SMF), whose output facet is fixed within the calibration module at a plane conjugate to the focal plane of the calibration objective (see Supplementary Figure \ref{suppl:fig:setup}). The phantom light propagates through the probe and the detection path, emerging at DMD2 as an annular intensity distribution. To visualise this ring, small DMD2 super-pixels ($6\times6$ mirrors) are sequentially configured to direct the reflected signal toward PMT1, effectively raster-scanning the light distribution at DMD2. Using this feedback, the plane producing the minimum-thickness annulus at DMD2 is selected.

Prior to imaging, the first step is to measure the transmission matrix of the excitation path, following the procedure described in \cite{Stiburek2023NC}. Once obtained, DMD1 is configured to generate the required light modulation for producing diffraction-limited foci across the confocal plane of the probe. Subsequently, calibration of the confocal detection path is performed, yielding the parameters of the virtual apertures (collection annuli) displayed on DMD2 for each excitation focus within the calibrated field of view (FOV). Due to experimental constraints, neither the distal confocal plane nor the DMD2 plane can be assumed perfectly isotropic; both may exhibit real or virtual ellipticity during the calibration.
To establish a mapping between the excitation focus positions and the corresponding confocal mask parameters at DMD2, we therefore introduce nine free parameters: the 2D centre position, ellipticity (eccentricity), and ellipse orientation for each plane, linked by a single global scaling factor.
These parameters are estimated from measurements performed at approximately 100 output points distributed across the FOV of the confocal fibre probe. During the acquisition, the entire calibration module is sequentially translated to different positions. At each position, the fluorescent phantom is moved to a distinct location within the fibre focal plane, producing a corresponding annular distribution at DMD2. The ring-shaped intensity distribution at DMD2 is acquired once again, by sequentially redirecting $6\times6$ DMD2 super-pixels toward PMT1. In parallel, the position of the fluorescent phantom is independently determined by scanning the excitation foci across the FOV using DMD1 and recording the signal on the photodiode positioned behind the SMF. Only excitation foci, which overlap with the fluorescent phantom, are coupled into the SMF and produce a detectable signal. Since both these procedures rely on different DMDs and detectors, they are performed simultaneously.
Finally, a model function is fitted to the complete dataset, yielding all required parameters and enabling the generation of appropriate confocal masks on DMD2 for all excitation points within the FOV. Notably, multiple excitation points lying along an ellipse in the FOV -- each requiring a unique DMD1 pattern -- share a single confocal mask on DMD2. As a result, while the full onboard storage capacity of DMD1 is typically utilised, the storage requirements for DMD2 are substantially reduced. Consequently, when multiple filtration levels (characterised by different $w$ factors) are required, several sets of confocal masks with varying annulus thicknesses can be computed in advance and uploaded to DMD2 for flexible $w$-factor selection.

The $w$-factor is in this work expressed relatively to $w_{floor}$ which is derived from the diffraction limit of the system.  
To understand this quantity, consider a Gaussian beam at the wavelength of \SI{520}{\nano\metre} with a flat wavefront at the plane of its waist, which aligns with the proximal facet of the fibre. 
The waist size of the beam matches the core size of the SI-part of the fibre probe.   
The beam propagates towards DMD2 through lenses L8, L10 and L11 (see Supplementary Figure \ref{suppl:fig:setup}). When it reaches the plane of the DMD2, its size (the diameter of the area over which the beam's amplitude exceeds 1/e of its peak value) is 3$\times$ $w_{floor}$. 

\subsection{PSF measurement}

To comprehensively describe the confocal confocal point spread function (PSF), we measured the complete
three-dimensional excitation and collection PSFs. The
SMF within the calibration module
was employed to sample the excitation PSF and further served as a point source
for the simultaneous acquisition of the collection PSF. 
The 3-D measurement sequence involved laterally moving the
end of the SMF in 2-D with respect to the calibration module, thereby positioning the virtual fluorescence phantom across the focal plane of the calibration objective. 
Scanning the PSFs along the third (axial) dimension was 
achieved by displacing the whole calibration module with respect to the confocal probe. 
While executing the 3-D raster scan across the volume behind the confocal probe's distal end we acquired the intensity of the excitation beam coupled into the SMF
 on the photodiode (PD), providing the excitation point spread function, Figure~\ref{fig:psf}a.
Simultaneously, the intensity of the fluorescent phantom (light from the green LED delivered by SMF and the calibration module optics towards confocal probe), filtered by DMD2 virtual aperture was measured on PMT1, providing the collection point
spread function, Figure~\ref{fig:psf}b. The effective PSF
(Figure~\ref{fig:psf}c) was then computed as the product of the excitation and
the collection PSFs. 
Exploiting the high refresh rate of the DMD technology, allowed us to rapidly change the patterns displayed on
both DMDs during single setting of the translational stages. This facilitated the measurement of
excitation and collection PSFs for multiple positions of the focal point and
various widths of the confocal masks. 

\subsection{Heartbeat compensation}
\label{met:HB compensation}

Acquisition of single full-resolution and full FOV frame is typically completed in a few seconds (\SI{2.5}{\second} for Fig. \ref{fig:imaging} {\bf b}--{\bf e}).
With typical heartbeat frequencies of a mouse ($\approx \SI{10}{\hertz}$), we can expect several 10s of heartbeat pulses quasi-periodically shifting the sample scene. The amplitude of the motion ranges from fractions to units of \si{\micro\metre}, depending on the proximity, concentration and size of blood vessels but also the proximity to skull and coverglass surfaces, which represent alleviating factors.   
Raster-scanning the sample row-by-row will thereby introduce several 10s of virtual tissue stretches and compressions (as seen Fig. \ref{fig:imaging} {\bf b}--{\bf c}).  

Assuming the heartbeat is periodic, we can define its phase as a normalised measure of time within the cardiac cycle, which facilitates the alignment of motion artefacts across heartbeats. As shown in Supplementary Movie \ref{hb_expl}, the heartbeat phase changes rapidly across the image rows, producing a virtual deformation of structures with relatively high spatial frequency. Because many heartbeat pulses occur during a single-frame acquisition, the phase values are distributed nearly uniformly across pixels (e.g., a similar number of pixels fall within each decile of the phase interval). Given the frame rate of our imaging system and the heartbeat frequency, we inevitably acquire some pixels at the beginning, middle, and end of each cardiac cycle. However, the random-access capability of our imaging modality -- allowing pixels to be acquired in any order -- enables us to choose when each pixel is recorded.

In this regard, we can define alternative scanning orders, which would have more favourable phase distributions. Particularly for the purpose of observing details of structural connectivity and identifying their changes in time, we would benefit if the spatial frequencies of the virtual tissue deformation were as low as possible. The optimum phase distribution would stretch the opposite extremities of the phase interval across the pixels as far as possible (e.g., top-left and bottom-right corner of the image) and smoothly fill the pixels in-between, leading to minimum phase difference across the neighbouring pixels. Combining this with the prediction of the instantaneous heartbeat phase throughout the scan, we can determine an acquisition order that best matches this prescribed phase map. 
This approach is, however, very sensitive to the accuracy of the heart-beat frequency estimation. When the heartbeat frequency start deviating from the estimate, progressively worsening dithering  artefacts follow rapidly, making this method impractical for our case.

An alternative way is to scan the sample block-wise. 
Each block comprises a specified number of rows, selected such that the scan time for a single block matches, as closely as possible, the duration of one heartbeat. The blocks are then scanned column by column keeping the phase difference amongst pixels still relatively small, and aligning the phase of each block well with the previous. As seen in Supplementary Movie \ref{hb_expl}, the approach is much less sensitive to the accuracy of the heart-beat frequency estimation. Fig. \ref{fig:imaging} compares this scanning sequence ({\bf d, e}) to the standard raster ({\bf b, c}). 
All above-mentioned scanning sequences are assessed in Supplementary Figure \ref{HB_scale} -- a simulation with timing and scale parameters closely approaching our experimental conditions. 

For our volumetric datasets, we use a modified block-wise scanning sequence. Within the coordinate system shown in Supplementary Movie \ref{hb_expl}, we form a block in the $x-z$ plane (column - layer). Typically we only use a single block matching the single heartbeat period requirement. The scanning of the block is organised layer-wise.

\subsection{Animals}

All animal experiments were conducted in accordance with protocols approved by
the Branch Commission for Animal Welfare of the Ministry of Agriculture of the
Czech Republic (permissions no. 55/2018, 47/2020, 49/2020 and 29-2023-P). All transgenic
lines were obtained from Jackson Laboratory, Maine, USA. The Thy1-GFP line M
(Stock No: 007788), Thy1-GFP line H (Stock No: 003782) were used for imaging of neurons, line CX3CR-1\textsuperscript{GFP} (Stock No:005582) for imaging of microglia.
Floxed Ai162D (Stock No: 031562) crossed
with cre-expressing CaMK2a-CreERT2 (StockNo: 012362) treated at the age of
4~weeks with Tamoxifen (\qty{75}{\mg/\kg} of body weight in a single dose,
Sigma-Aldrich) was used for calcium imaging. Both male and female mice aged
\numrange{20}{84} weeks were used. Mice were housed in a \qty{12}{\hour}
light/\qty{12}{\hour} dark cycle at \qtyrange{21}{23}{\degreeCelsius} and
\qtyrange{50}{60}{\percent} relative humidity with ad libitum access to food and
water.

\subsection{Surgical procedures}
\label{met:Surgery}

The mouse was anaesthetised by inhalation of isoflurane (\qty{5}{\percent} for induction, \qty{1.5}{\percent} for
maintenance) in pure \ce{O2} and placed on a portable platform. A warming
blanket with a feedback loop from a rectal probe was maintaining the
body temperature at \qty{37}{\degreeCelsius}. An i.p. injection of dexamethasone
(\qty{0.125}{\ml}, \qty{2}{\mg/\kg}) and buprenorphine (\qty{0.042}{\ml}, \qty{0.05}{\mg/\kg})
were administered to reduce inflamation and edema, and for analgesia, respectively. Further, atropine (\qty{0.063}{\ml}, \qty{0.05}{\mg/\kg}) and
gluose (\qty{0.1}{\ml}, \qty{400}{\mg/\kg}) were administered s.c. to ease breathing and for hydration, respectively. After depilation of the fur and 
local application of lidocaine 
( \qty{2}{\percent}, \qty{0.05}{\ml}), skin over the target area was
removed by a round incision. The $xy$ coordinates of the target area were
identified using the mouse brain atlas \cite{Paxinos2019} and the position of
bregma and the target area were marked on the surface of the skull. A custom
head plate was attached by a cyanoacrylate gel and aditionally secured by dental cement 
(Jet Dental Repair Powder and Liquid, Lang Dental) to aid head-fixation of the animal during the imaging sessions. 
Finally, a high-speed drill (tip diameter \qty{0.3}{\mm}) was used to make a small round craniotomy centred over
the target area and the dura was removed. For acute experiments under anesthesia, the mouse was
transferred directly to the endoscope and imaged. At the end of the session, the mouse was sacrificed by anaesthesia overdose of ketamine/xylazine
(\qty{0.1}{\ml} of ketamine, \qty{200}{\mg/\kg}, and
\qty{0.026}{\mg} of xylazine, \qty{20}{\mg/\kg}) administered
i.p. and cervical dislocation.

For awake experiments, the procedure remains identical until the point of the
dura removal. The craniotomy is then filled with agar (\qty{1.5}{\percent} in
ACSF) and closed using a cover glass featuring 3~circular ports (\qty{300}{\um}
diameter each) to allow insertion of the endoscopic probe. The ports were
ablated into the cover glass by an IR laser. The exposed area around the cover
glass is filled with dental cement and protected with a
fast-curing silicon (Smooth-On). The anesthesia is discontinued and the animal
recovers in a warmed cage. The mouse is injected with 3~doses of ketoprofen
(\qty{0.05}{\ml}, \qty{5}{\mg/\kg}) s.c., first
dose at least one hour before the end of the surgery and 2~additional doses on
consecutive days. The mouse is habituated for handling and head-fixation for at
least 5~days. At the imaging session, the silicone is removed and the probe
enters the tissue through the port. In some sessions, a visual stimulus to the contralateral eye using blue LED flashes were used.
The animals were imaged up to 10 times according to the approved protocol. At the end of the experiment, the
mouse is sacrificed in the same manner as in the acute experiments.

\end{multicols}

\clearpage
\onecolumn
\section*{\Large Supplementary Information}

\renewcommand\thefigure{S\arabic{figure}}    
\renewcommand\thesection{S\arabic{section}}    
\renewcommand\thesubsection{S\arabic{section}.\arabic{subsection}}    
\renewcommand\theequation{S\arabic{equation}}

\setcounter{figure}{0}    
\setcounter{page}{1}    
\setcounter{equation}{0}

\subsection*{Supplementary Figures}
\begin{figure*}[!ht]
\centerline{
	\includegraphics[width=1\textwidth]{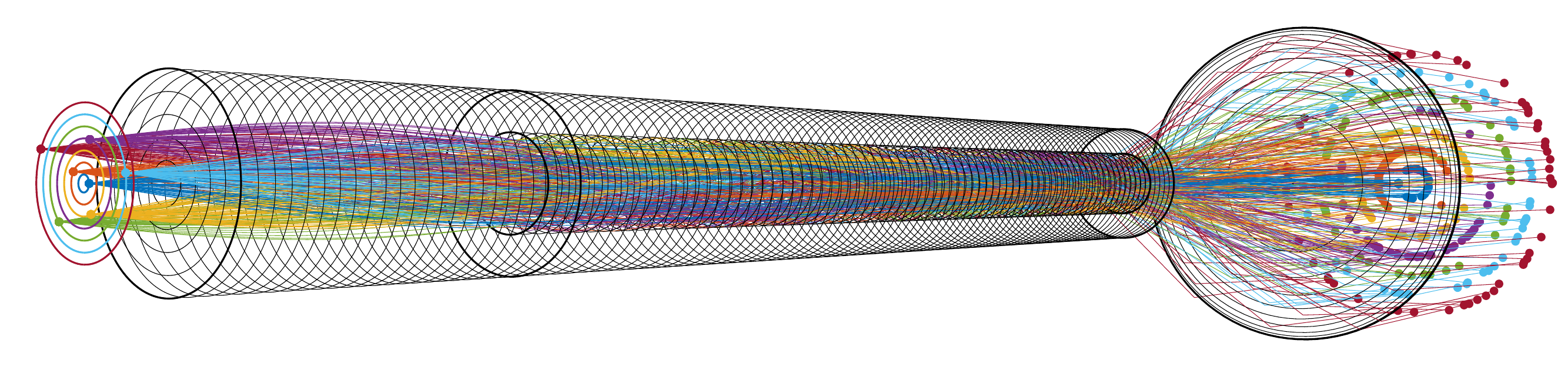}}
	\caption{
{\bf An illustrative ray-model of the light transport for the straight-view terminated composite MMF fibre probe.}
The medium outside the distal end of the probe (left) is considered to be water, while the medium outside the proximal end (right) is considered to be the air. 
Individual light paths originate outside the probe near its distal extremity in a series of foci. Each focus lies in the focal plane for which the length of the endcap has been optimised. Each focus is further located at a specific radial distance from the optical axis. The model traces the rays all the way to the proximal far-field plane, passing through the water-filled space, the straight-view-terminated GRIN endcap, the SI segment and the proximal lens surrounded by two air-filled volumes. 	
	}
	\label{ray_str}
\end{figure*}

\begin{figure*}[!ht]
\centerline{
	\includegraphics[width=1\textwidth]{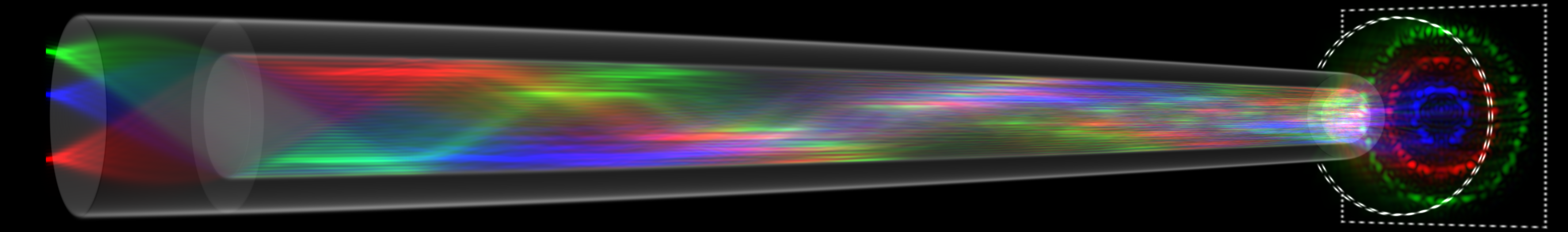}}
	\caption{{\bf Numerical simulation of light propagation throughout the composite fibre probe, considering single wavelength of $\lambda_{vac}=\SI{511}{\nano\metre}$}. The GRIN fibre forming the endcap has core of \SI{78}{\micro\metre} in diameter, on-axis refractive index of 1.45 and NA of 0.29. Light propagation in both fibre types is modelled as linear superposition of propagation invariant modes \cite{Ploschner2015NP,BoonzajerFlaes2018PRL}.
	The endcap length is set to \SI{248}{\micro\metre}, which leads to the working distance of \SI{50}{\micro\metre}. The SI MMF has core of \SI{50}{\micro\metre} in diameter and the NA of 0.22. The length of the SI segment is set to \SI{4}{\milli\metre}. The focal length of the focusing lens (indicated in the geometry by a pair of dashed lines) is \SI{500}{\micro\metre}. Each of the RGB channels represents intensity distribution originating in a focus at different radial position across the focal plane. The light distribution is `filtered' by the composite fibre, i.e., only light signal that is transported through the probe is visible. The imaged volume is compressed along the direction of the fibre axis by factor of 2. The image is computed in spherical projection, placing camera \SI{1.5}{\milli\metre} in front of the probe's distal facet and \SI{3}{\milli\metre} off the probe's axis. 
	}
	\label{wave_511}
\end{figure*}

\begin{figure*}[!ht]
\centerline{
	\includegraphics[width=1\textwidth]{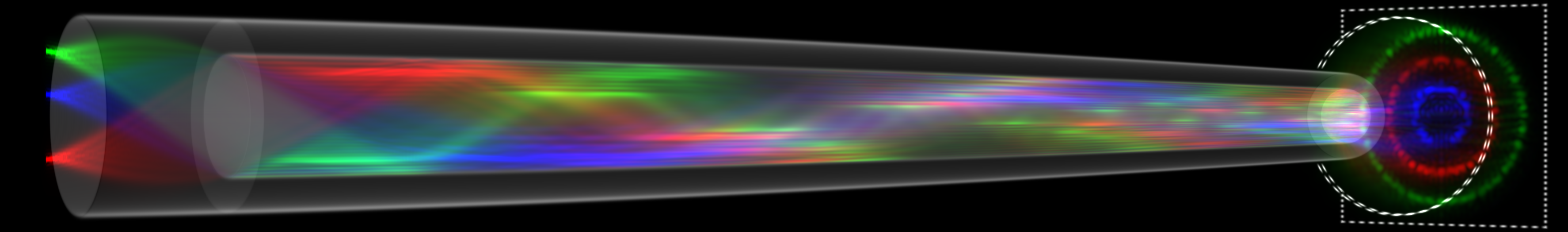}}
	\caption{{\bf Numerical simulation of light propagation throughout the composite fibre probe, considering the complete GFP spectrum.} Single wavelength intensities (as in Fig. \ref{wave_511}) were calculated from \SI{465}{\nano\metre} to \SI{651}{\nano\metre} with spectral step of \SI{2}{\nano\metre} (see Supplementary Movie \ref{wave_lambdas}) to be summed with weights given by EGFP emission spectrum. The composite fibre and the geometry of the image is identical to Fig. \ref{wave_511}. 
	}
	\label{wave_gfp}
\end{figure*}

\begin{figure*}[!ht]
\centerline{
	\includegraphics[width=.7\textwidth]{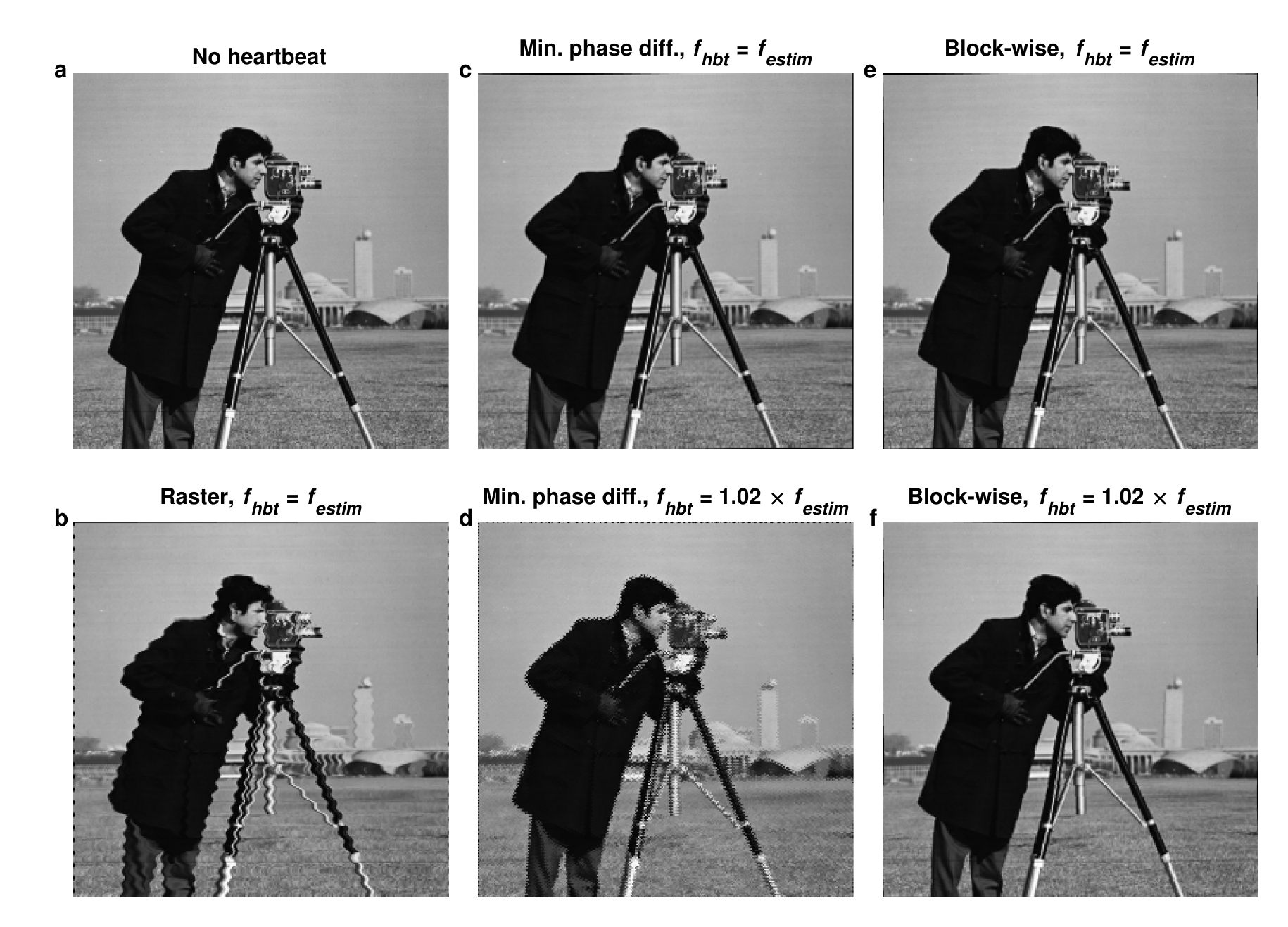}}
	\caption{{\bf Simulation of heartbeat-induced motion artefacts for various scanning sequences}. The chosen object is the standard grayscale test image ‘cameraman.tif’ (resolution 256 $\times$ 256) provided in MATLAB’s ``Image Processing Toolbox.” The heartbeat tissue motion is modelled circular, with amplitude of three pixels (corresponding to $\approx \SI{0.86}{\micro\metre}$). Its frequency is modelled at \SI{10}{\hertz} to match the estimated value of the imaging algorithms, or \SI{10.2}{\hertz} to stress-test the imaging algorithms on inaccurate frequency estimate. 
	{\bf a}, No heart-beat induced object motion (the object itself).  
	    {\bf b}, Image obtained with the standard, raster-scanning sequence (row by row).
	     {\bf c}, Image obtained with the scanning sequence for minimum heartbeat phase difference between pixels, heartbeat frequency estimate matches the true heartbeat frequency.
	     {\bf d}, Image obtained with the scanning sequence for minimum heartbeat phase difference between pixels, heartbeat frequency estimate deviates from the true heartbeat frequency by \SI{2}{\percent}.
	     {\bf e}, Image obtained with the block-wise scanning sequence, heartbeat frequency estimate matches the true heartbeat frequency.
	     {\bf f}, Image obtained with the block-wise scanning sequence, heartbeat frequency estimate deviates from the true heartbeat frequency by \SI{2}{\percent}.
	}
	\label{HB_scale}
\end{figure*}

\begin{figure*}[!ht]
\centerline{
	\includegraphics[width=.7\textwidth]{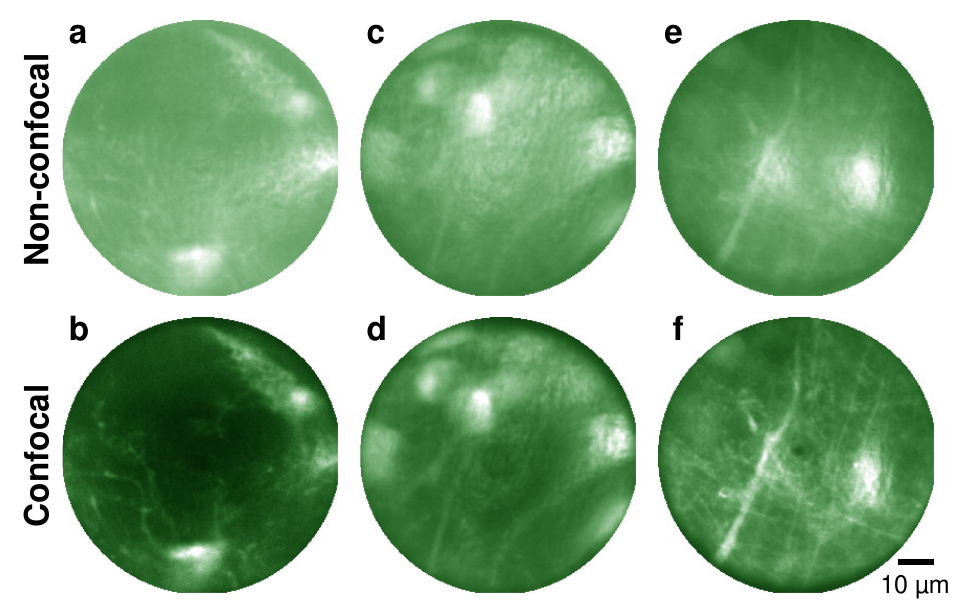}}
	\caption{{\bf Full-field of view non-confocal and confocal records in various specimens}. {\bf a} and {\bf b}, Microglia in a CX3CR-1\textsuperscript{GFP} mouse line. {\bf c} and {\bf d}, Neurones in a Thy1-GFP-M.   {\bf e} and {\bf f}, Neurones in a Thy1-YFP-H line. 
	}
	\label{full_field_other}
\end{figure*}

\begin{figure*}[!ht]
	\centerline{
	\includegraphics[width=\textwidth]{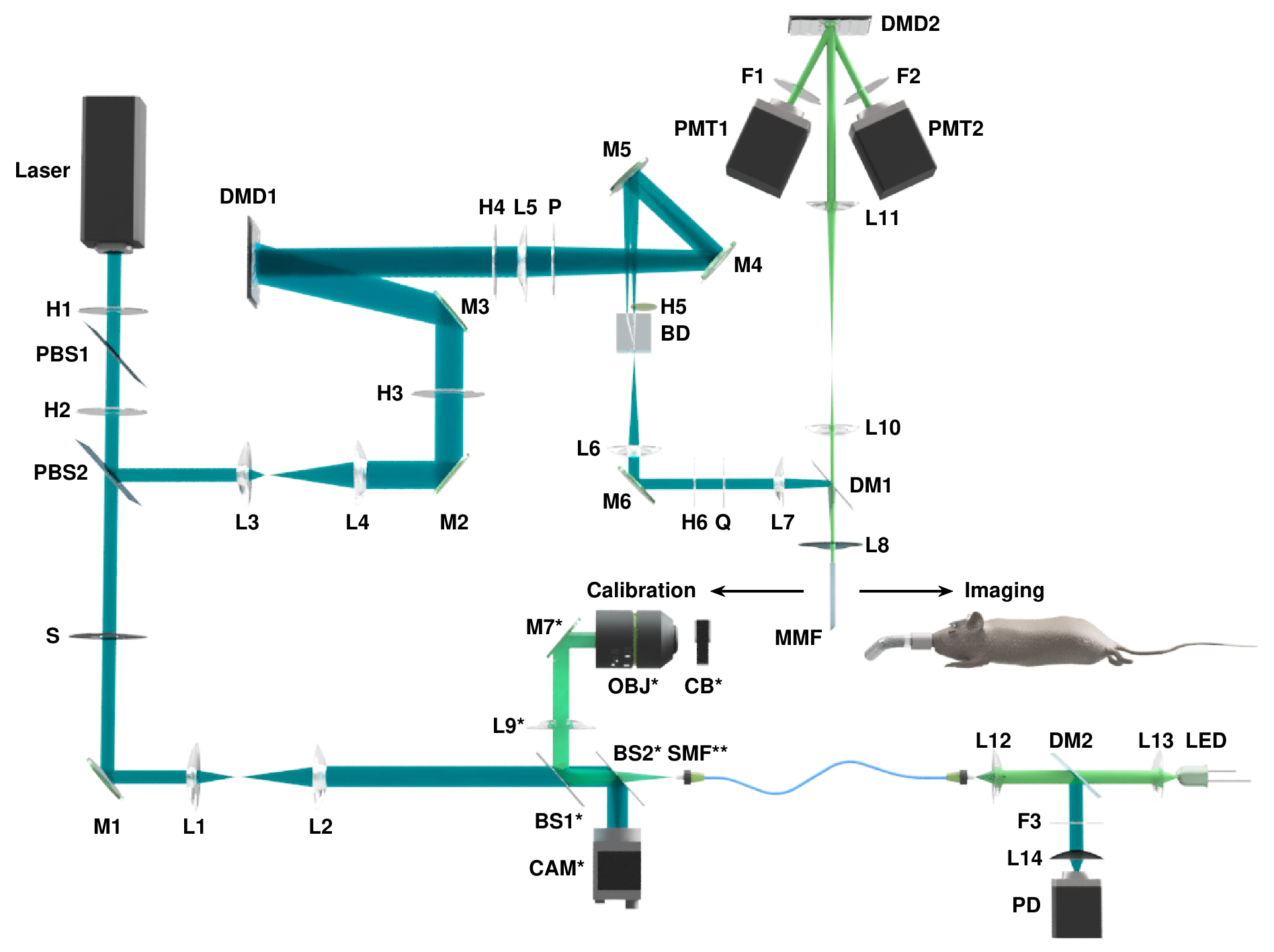}}
	\caption{
		{\bf The experimental geometry.}
		\newline ~~~ \newline
      \begin{minipage}[t]{0.48\textwidth}
        \raggedright
		\textbf{BD}: calcite beam displacer (Thorlabs BD40),
		\newline \textbf{BS1,2}: beamsplitter plate (Thorlabs BSW10),
		\newline \textbf{CAM}: camera (Basler ace acA640-750um),
		\newline \textbf{CB}: calibration basin,
		\newline \textbf{DM1}: dichroic mirror (Chroma T495lpxru),
		\newline \textbf{DM2}: dichroic mirror (Chroma T495lpxru),
		\newline \textbf{DMD1,2}: digital micromirror device (ViALUX V-7001),
		\newline \textbf{F1,2}: bandpass filter (Chroma ET510/20m),
		\newline \textbf{F3}: laser line filter (Thorlabs FL488-1),
		\newline \textbf{H1--4}: half-wave plate (Edmund Optics \#46-551),
		\newline \textbf{H5}: half-wave plate (Newport 10RP02-12),
		\newline \textbf{H6}: half-wave plate (Edmund Optics \#46-551),
		\newline \textbf{L1}: lens (Thorlabs C240TMD-A, $f=\qty{8}{mm}$),
		\newline \textbf{L2}: lens (Thorlabs AC254-300-A-ML, $f=\qty{300}{mm}$),
		\newline \textbf{L3}: lens (Thorlabs C240TMD-A, $f=\qty{8}{mm}$),
		\newline \textbf{L4}: lens (Thorlabs AC254-125-A-ML, $f=\qty{125}{mm}$),
		\newline \textbf{L5}: lens (Thorlabs AC254-300-A-ML, $f=\qty{300}{mm}$),
		\newline \textbf{L6}: lens (Thorlabs AC254-75-A-ML, $f=\qty{75}{mm}$),
		\newline \textbf{L7}: lens (Thorlabs AC254-150-A-ML, $f=\qty{150}{mm}$),
		\newline \textbf{L8}: lens (Thorlabs C240TMD-A, $f=\qty{8}{mm}$),
	     \end{minipage}
	         \begin{minipage}[t]{0.48\textwidth}
	           \raggedright
		\textbf{L9}: lens (Thorlabs AC254-150-A-ML, $f=\qty{150}{mm}$),
		\newline\textbf{L10}: lens (Thorlabs AC254-50-A-ML, $f=\qty{50}{mm}$),
		\newline\textbf{L11}: lens (Edmund Optics \#47-644, $f=\qty{175}{mm}$),
		\newline\textbf{L12}: lens (Thorlabs F110APC-532, $f=\qty{6.09}{mm}$),
		\newline\textbf{L13}: lens (Thorlabs AC254-30-A-ML, $f=\qty{30}{mm}$),
		\newline\textbf{L14}: lens (Thorlabs AC254-30-A-ML, $f=\qty{30}{mm}$),
		\newline\textbf{Laser}: laser (Coherent Sapphire 488 SF NX),
		\newline\textbf{LED}: green LED (GeTian GT-P03G6410160),
		\newline\textbf{M1--7}: dielectric mirror (Thorlabs BB1-E02),
		\newline\textbf{MMF}: multi-mode fibre-based probe,
		\newline\textbf{OBJ}: objective (Olympus 20X X-Apo 0.80NA/0.6WD),
		\newline\textbf{PBS1,2}: polarising beamsplitter (Thorlabs CCM1-PBS251/M),
		\newline\textbf{PD}: photodetector (Thorlabs PDA36A2),
		\newline\textbf{PMT1,2}: amplified GaAsP photomultiplier tube (Thorlabs
			PMT2101/M),
		\newline\textbf{POL}: polariser (Thorlabs LPVISA100-MP2),
		\newline\textbf{S}: shutter (Thorlabs SHB025T),
		\newline\textbf{SMF}: single mode fibre (Thorlabs P3-488PM-FC-1),
		\newline\textbf{Q}: quarter-wave plate (Thorlabs WPMQ10M-488).   
	            \end{minipage}
	            ~~~\newline~~~
	            \newline$^{\star}$ The marked components form the calibration unit. They are rigidly anchored to a single motorised XYZ translation stage composed of Physik Instrumente L-511.2ASD00 and Physik Instrumente L-731.44SD. 
		\newline
		$^{\star\star}$ The marked end of {\bf SMF} is housed on a separate XY motorized translation stage.
	}
	\label{suppl:fig:setup}
\end{figure*}

\clearpage
\renewcommand\thefigure{SM\arabic{figure}}  
\setcounter{figure}{0}      
\subsection*{Supplementary Movies}
\begin{figure*}[!ht]
\centerline{
	\includegraphics[width=1\textwidth]{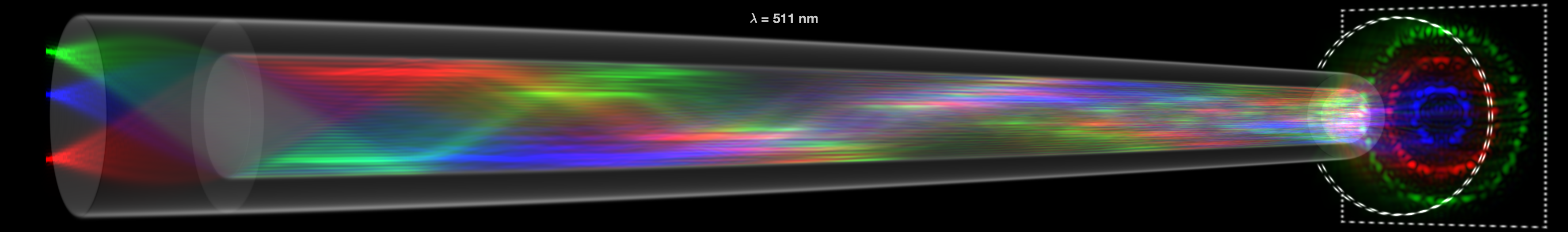}}
	\caption{{\bf Numerical simulation of light propagation throughout the composite fibre probe, for individual wavelengths within the EGFP emission spectrum}.  The composite fibre and the geometry of the image is identical to Fig. \ref{wave_511}. Media available from \href{https://zenodo.org/records/18017907/files/SM1.mp4?download=1}{zenodo.org}.
	}
	\label{wave_lambdas}
\end{figure*}

\begin{figure*}[!ht]
\centerline{
	\includegraphics[width=1\textwidth]{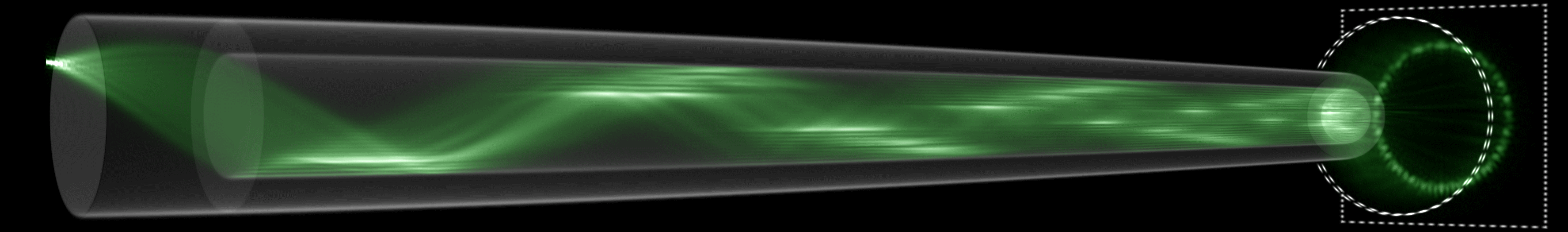}}
	\caption{{\bf Numerical simulation of light propagation throughout the composite fibre probe, considering complete EGFP emission spectrum}. Within the animation, the focal point varies its distance from the axis. The composite fibre and the geometry of the image is identical to Fig. \ref{wave_511}. Media available from \href{https://zenodo.org/records/18017907/files/SM2.gif?download=1}{zenodo.org}.
	}
	\label{wave_mov1d}
\end{figure*}

\begin{figure*}[!ht]
\centerline{
	\includegraphics[width=1\textwidth]{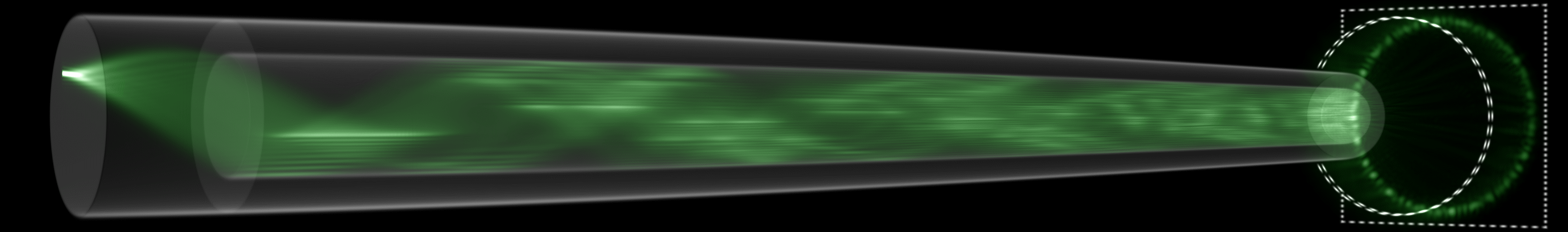}}
	\caption{{\bf Numerical simulation of light propagation throughout the composite fibre probe, considering complete EGFP emission spectrum}. Within the animation, the focal point revolves aroud the axis. The composite fibre and the geometry of the image is identical to Fig. \ref{wave_511}. Media available from \href{https://zenodo.org/records/18017907/files/SM3.gif?download=1}{zenodo.org}.
	}
	\label{wave_movcirc}
\end{figure*}

\begin{figure*}[!ht]
  \centering
  \begin{minipage}[t]{0.48\textwidth}
    \centering
    \includegraphics[width=\linewidth]{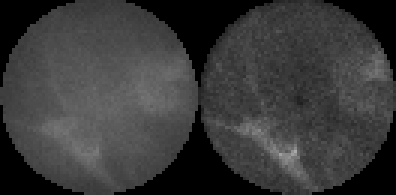}
    \caption{{\bf Non-confocal (left) and confocal (right) records of signalling activity of somata in hippocampus.} Media available from \href{https://zenodo.org/records/18017907/files/SM4.mp4?download=1}{zenodo.org}.}
    \label{sig_soma}
  \end{minipage}\hfill
  \begin{minipage}[t]{0.48\textwidth}
    \centering
    \includegraphics[width=\linewidth]{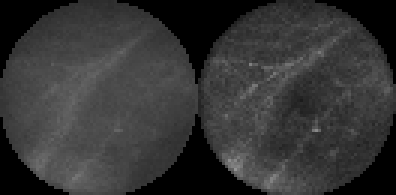}
    \caption{{\bf Non-confocal (left) and confocal (right) records of signalling activity of dendrites in hippocampus.} Media available from \href{https://zenodo.org/records/18017907/files/SM5.mp4?download=1}{zenodo.org}.}
    \label{sig_proc}
  \end{minipage}
\end{figure*}

\begin{figure*}[!ht]
\centerline{
	\includegraphics[width=1\textwidth]{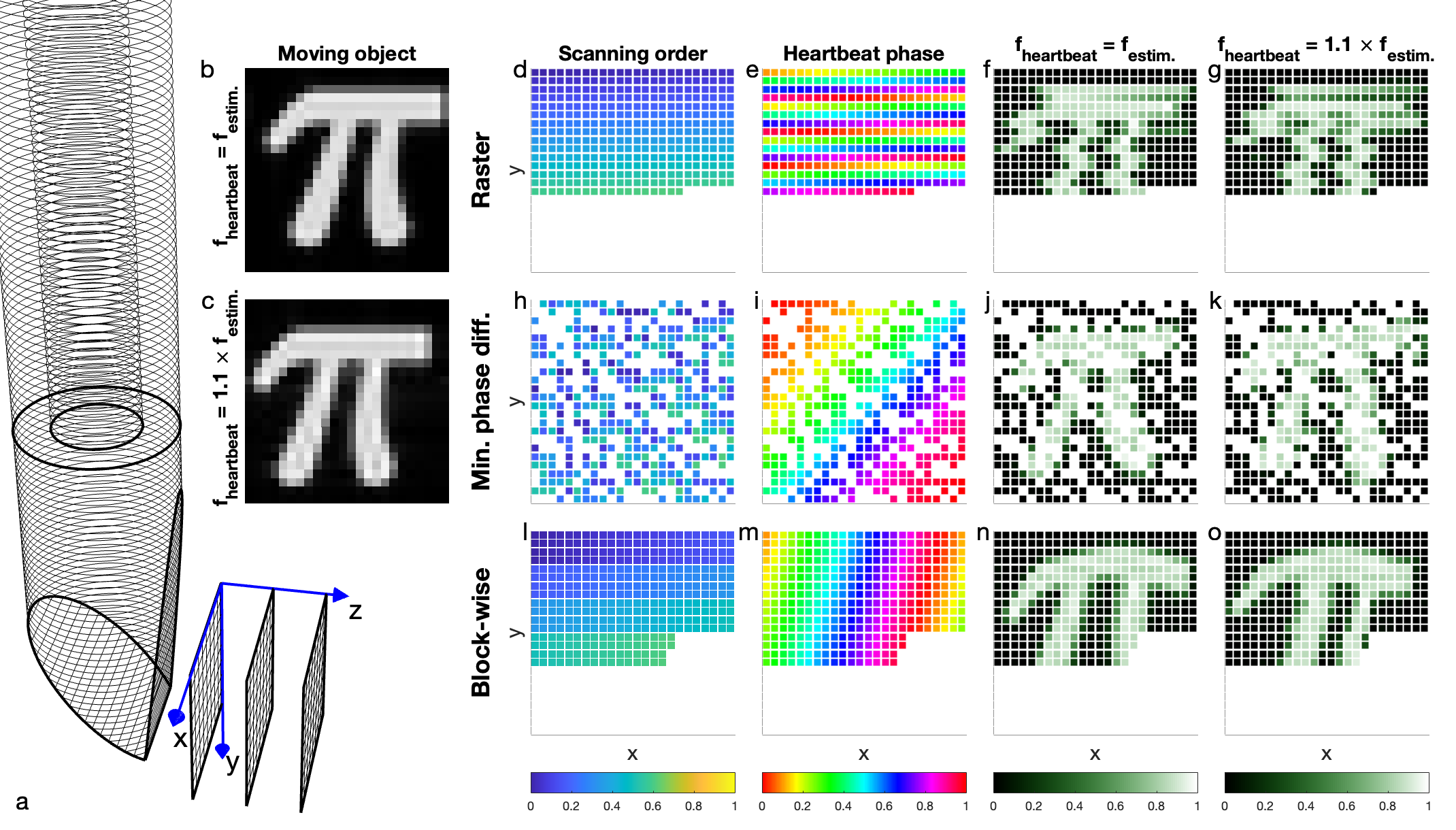}}
	\caption{{\bf Principle of heartbeat-induced motion artefact suppression.} 
	{\bf a}, The coordinate reference frame. 
	{\bf b}, The object subjected to periodic heartbeat-induced motion. The motion frequency matches the estimated value.  
	{\bf c}, The object subjected to periodic heartbeat-induced motion. The motion frequency deviates from the estimated value.  
	{\bf d} -- {\bf g}, The scanning order, the heartbeat phase, the resulting image obtained for object {\bf b} and the resulting image obtained for object {\bf c}, respectively, for the standard raster-scanning sequence. 
	{\bf h} -- {\bf k}, The equivalent of {\bf d} -- {\bf g} for the scanning sequence demanding minimum phase difference between neighbouring pixels. 
	{\bf l} -- {\bf o}, The equivalent of {\bf d} -- {\bf g} for the block-wise scanning sequence. Media available from \href{https://zenodo.org/records/18017907/files/SM6.mp4?download=1}{zenodo.org}.
	}
	\label{hb_expl}
\end{figure*}

\begin{figure*}[!ht]
\centerline{
	\includegraphics[width=.5\textwidth]{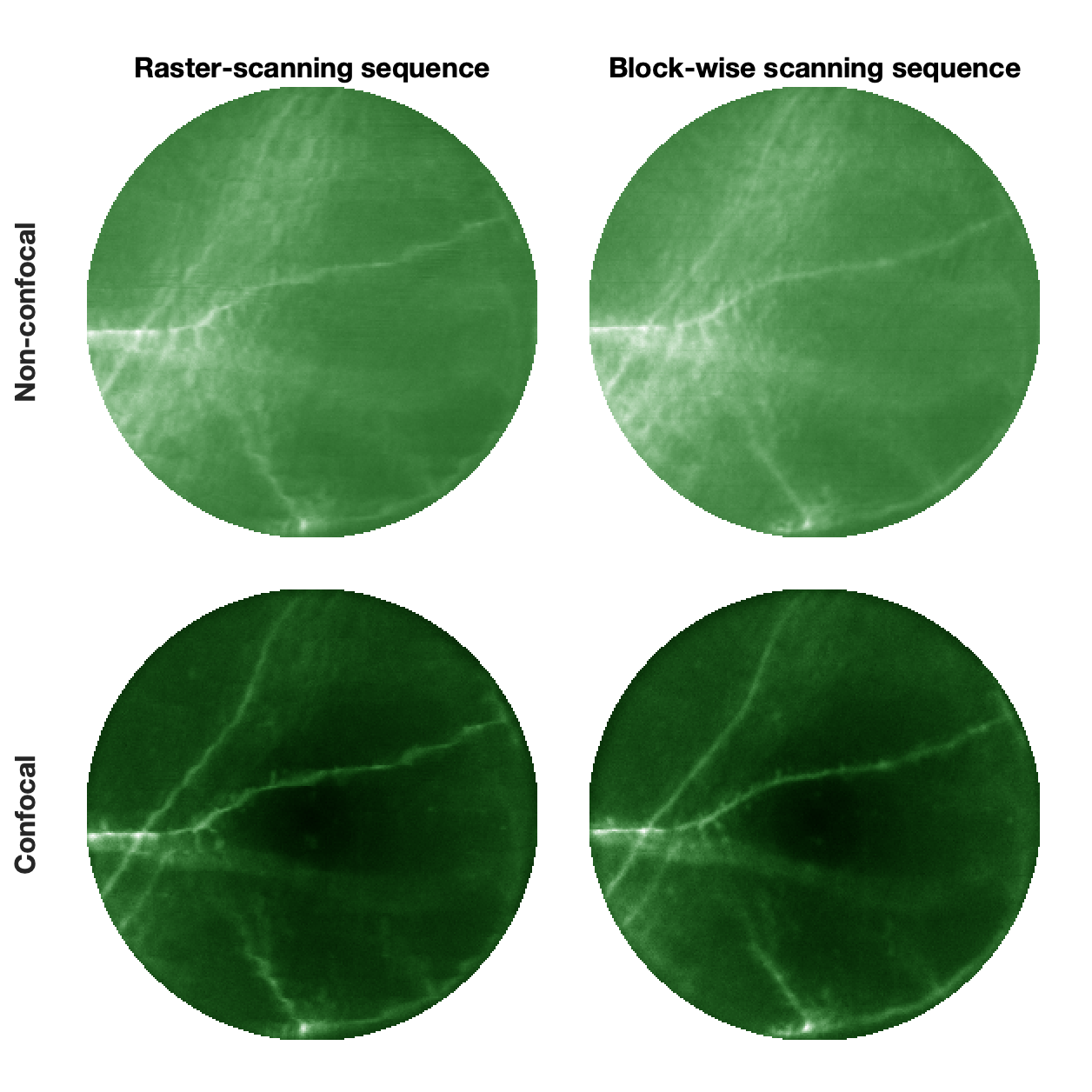}}
	\caption{{\bf Heartbeat-induced motion artefact suppression.} 
	The animation compares the results of the standard raster scanning sequence with the block-wise alternative for both non-confocal and confocal modalities, in 5 consecutive frames. Media available from \href{https://zenodo.org/records/18017907/files/SM7.gif?download=1}{zenodo.org}.
	}
	\label{hb_demo}
\end{figure*}

\begin{figure*}[!ht]
\centerline{
	\includegraphics[width=1\textwidth]{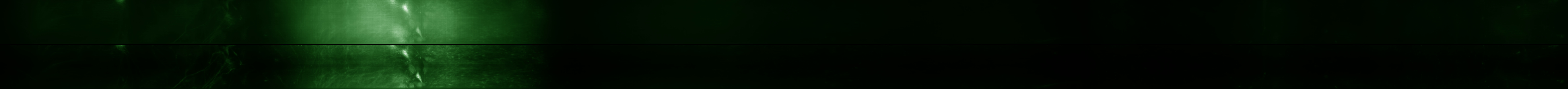}}
	\caption{{\bf Volumetric atlas-view of non-confocal (top) and confocal (bottom) records.} Original brightness, contrast and gamma. The animations shows different focal planes sequentially. Media available from \href{https://zenodo.org/records/18017907/files/SM8.gif?download=1}{zenodo.org}.
	}
	\label{atlas_orig_gif}
\end{figure*}

\begin{figure*}[!ht]
\centerline{
	\includegraphics[width=1\textwidth]{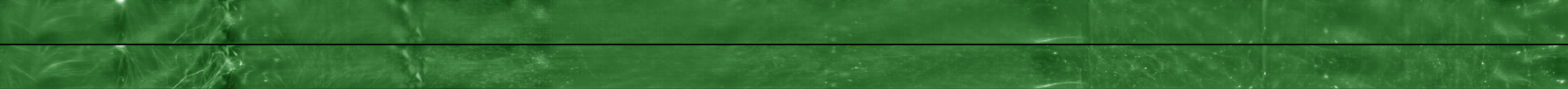}}
	\caption{{\bf Volumetric atlas-view of non-confocal (top) and confocal (bottom) records.} Brightness is locally normalised, contrast enhancement and gamma are applied globally. The animations shows different focal planes sequentially.   Media available from \href{https://zenodo.org/records/18017907/files/SM9.gif?download=1}{zenodo.org}.
	}
	\label{atlas_enh_gif}
\end{figure*}

\begin{figure*}[!ht]
\centerline{
	\includegraphics[width=1\textwidth]{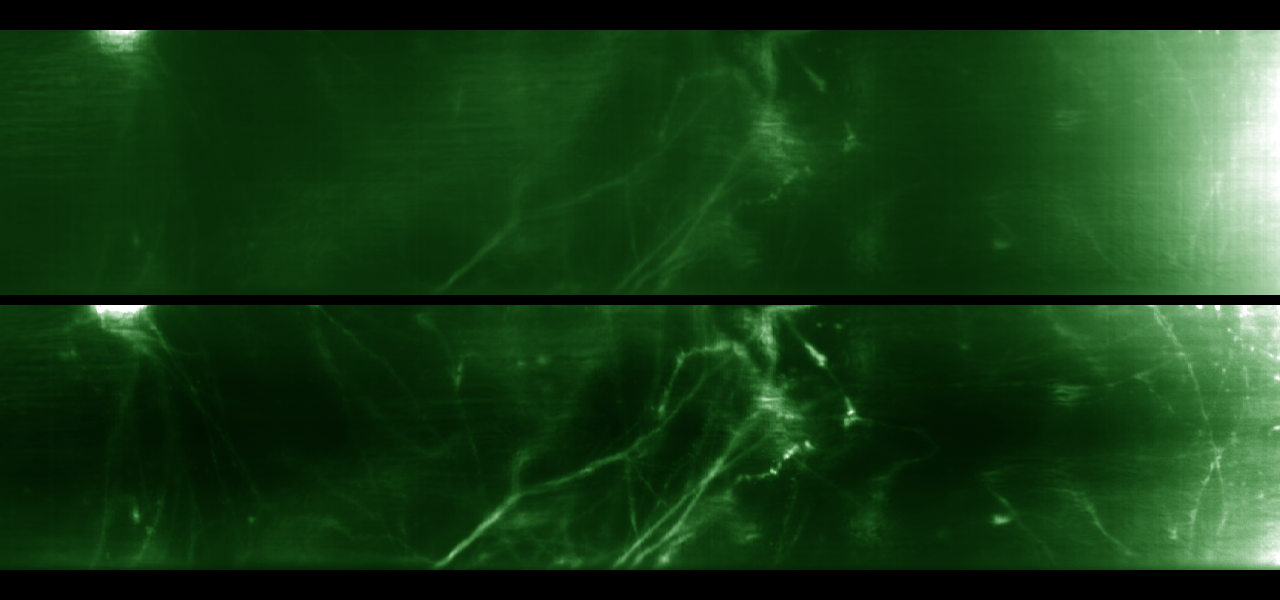}}
	\caption{{\bf Volumetric atlas-view of non-confocal (top) and confocal (bottom) records.} Brightness is adjusted such that \SI{0.1}{\percent} of the brightest pixels in each frame is saturated, no contrast and gamma manipulation. Movie focuses on different parts of the record.   Media available from \href{https://zenodo.org/records/18017907/files/SM10.mp4?download=1}{zenodo.org}.
	}
	\label{atlas_orig_mp4}
\end{figure*}

\begin{figure*}[!ht]
\centerline{
	\includegraphics[width=1\textwidth]{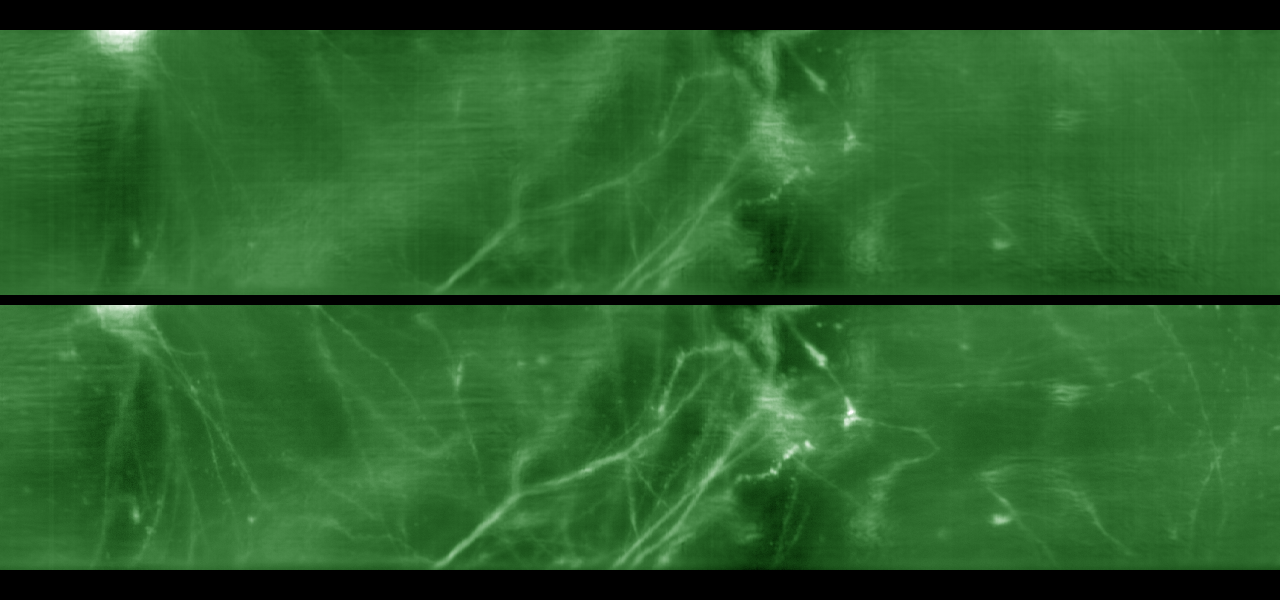}}
	\caption{{\bf Volumetric atlas-view of non-confocal (top) and confocal (bottom) records.} Brightness is locally normalised, contrast enhancement and gamma applied globally. Movie focuses on different parts of the record.   Media available from \href{https://zenodo.org/records/18017907/files/SM11.mp4?download=1}{zenodo.org}.
	}
	\label{atlas_enh_mp4}
\end{figure*}

\end{document}